\documentclass[journal]{IEEEtran}

\usepackage[T1]{fontenc}
\usepackage{amsmath,amssymb,amsfonts,mathrsfs,bm}% Typical maths resource packages
\usepackage{amsthm}
\usepackage{cite}
\usepackage{array}

\usepackage[shortlabels]{enumitem}
\usepackage{graphicx}
\usepackage{url}
\usepackage{color}
\usepackage{algorithm,algorithmic}
\usepackage{multirow}
\usepackage[table]{xcolor}
\usepackage[normalem]{ulem}
\usepackage{array}
\newcolumntype{L}[1]{>{\raggedright\let\newline\\\arraybackslash\hspace{0pt}}m{#1}}
\newcolumntype{C}[1]{>{\centering\let\newline\\\arraybackslash\hspace{0pt}}m{#1}}
\newcolumntype{R}[1]{>{\raggedleft\let\newline\\\arraybackslash\hspace{0pt}}m{#1}}
\usepackage{xparse}

\usepackage[hidelinks]{hyperref} %load as the last package

\ifx\useTheoremCounter\undefined
\newtheorem{Theorem}{Theorem}
\newtheorem{Corollary}{Corollary}
\newtheorem{Proposition}{Proposition}
\newtheorem{Lemma}{Lemma}
\else
\newtheorem{Theorem}{Theorem}

\newtheorem{Proposition}[Theorem]{Proposition}

\fi

\newtheorem{Definition}{Definition}
\newtheorem{Example}{Example}

\newtheorem{Assumption}{Assumption}

\newtheorem{Lemma_A}{Lemma}[section]
\newtheorem{Definition_A}{Definition}[section]

\theoremstyle{remark}

\newcommand{\Real}{\mathbb{R}}

\newcommand{\calD}{\mathcal{D}}

\newcommand{\calG}{\mathcal{G}}
\newcommand{\calH}{\mathcal{H}}
\newcommand{\calI}{\mathcal{I}}

\newcommand{\calO}{\mathcal{O}}

\newcommand{\calQ}{\mathcal{Q}}

\newcommand{\calS}{\mathcal{S}}

\newcommand{\calX}{\mathcal{X}}

\newcommand{\calZ}{\mathcal{Z}}

\newcommand{\bx}{\mathbf{x}}

\newcommand{\bz}{\mathbf{z}}

\newcommand{\scH}{\mathscr{H}}

\DeclareSymbolFont{bsfletters}{OT1}{cmss}{bx}{n}
\DeclareSymbolFont{ssfletters}{OT1}{cmss}{m}{n}
\DeclareMathSymbol{\bsfGamma}{0}{bsfletters}{'000}
\DeclareMathSymbol{\ssfGamma}{0}{ssfletters}{'000}
\DeclareMathSymbol{\bsfDelta}{0}{bsfletters}{'001}
\DeclareMathSymbol{\ssfDelta}{0}{ssfletters}{'001}
\DeclareMathSymbol{\bsfTheta}{0}{bsfletters}{'002}
\DeclareMathSymbol{\ssfTheta}{0}{ssfletters}{'002}
\DeclareMathSymbol{\bsfLambda}{0}{bsfletters}{'003}
\DeclareMathSymbol{\ssfLambda}{0}{ssfletters}{'003}
\DeclareMathSymbol{\bsfXi}{0}{bsfletters}{'004}
\DeclareMathSymbol{\ssfXi}{0}{ssfletters}{'004}
\DeclareMathSymbol{\bsfPi}{0}{bsfletters}{'005}
\DeclareMathSymbol{\ssfPi}{0}{ssfletters}{'005}
\DeclareMathSymbol{\bsfSigma}{0}{bsfletters}{'006}
\DeclareMathSymbol{\ssfSigma}{0}{ssfletters}{'006}
\DeclareMathSymbol{\bsfUpsilon}{0}{bsfletters}{'007}
\DeclareMathSymbol{\ssfUpsilon}{0}{ssfletters}{'007}
\DeclareMathSymbol{\bsfPhi}{0}{bsfletters}{'010}
\DeclareMathSymbol{\ssfPhi}{0}{ssfletters}{'010}
\DeclareMathSymbol{\bsfPsi}{0}{bsfletters}{'011}
\DeclareMathSymbol{\ssfPsi}{0}{ssfletters}{'011}
\DeclareMathSymbol{\bsfOmega}{0}{bsfletters}{'012}
\DeclareMathSymbol{\ssfOmega}{0}{ssfletters}{'012}

\newcommand{\balpha}{\bm{\alpha}}
\newcommand{\bbeta}{\bm{\beta}}

\DeclareMathOperator*{\argmax}{arg\,max}
\DeclareMathOperator*{\argmin}{arg\,min}

\DeclareMathOperator*{\arginf}{arg\,inf}

\newcommand{\qednew}{\nobreak \ifvmode \relax \else
      \ifdim\lastskip<1.5em \hskip-\lastskip
      \hskip1.5em plus0em minus0.5em \fi \nobreak
      \vrule height0.75em width0.5em depth0.25em\fi}

\newcommand{\indicator}[1]{{\bf 1}_{\{{#1}\}}}

\newcommand{\ofrac}[1]{{\frac{1}{#1}}}

\newcommand{\ip}[2]{{\left\langle{#1},{#2}\right\rangle}}
\newcommand{\norm}[1]{{\left\lVert{#1}\right\rVert}}

\newcommand{\cond}[2]{\left. {#1}\, \middle| \, {#2} \right.}

\DeclareDocumentCommand \P { g d() g } {%
	\IfNoValueTF {#3}
	{%
		\IfNoValueTF {#1}
		{%
			\IfNoValueTF {#2}
			{%
				\mathbb{P}%
			}%
			{%
				\mathbb{P}\left({#2}\right)%
			}%
		}%
		{%
			\IfNoValueTF {#2}
			{%
				\mathbb{P}_{#1}%
			}%
			{%
				\mathbb{P}_{#1}\left({#2}\right)%
			}%		
		}%
	}%
	{%
		\IfNoValueTF {#1}
		{%
			\mathbb{P}\left(\cond{#2}{#3}\right)%
		}%
		{%
			\mathbb{P}_{#1}\left(\cond{#2}{#3}\right)%
		}%	
	}%
}

\DeclareDocumentCommand \E { g o g } {%
	\IfNoValueTF {#3}
	{%
		\IfNoValueTF {#1}
		{%
			\IfNoValueTF {#2}
			{%
				\mathbb{E}%
			}%
			{%
				\mathbb{E}\left[{#2}\right]%
			}%
		}%
		{%
			\IfNoValueTF {#2}
			{%
				\mathbb{E}_{#1}%
			}%
			{%
				\mathbb{E}_{#1}\left[{#2}\right]%
			}%		
		}%
	}%
	{%
		\IfNoValueTF {#1}
		{%
			\mathbb{E}\left[\cond{#2}{#3}\right]%
		}%
		{%
			\mathbb{E}_{#1}\left[\cond{#2}{#3}\right]%
		}%	
	}%
}

\definecolor{gray90}{gray}{0.9}

\newcommand{\hide}[1]{}
\graphicspath{{./Figures/}}

\usepackage{epstopdf}
\usepackage{empheq}
\DeclareGraphicsExtensions{.eps}

\newcommand{\hG}{\widehat{G}}
\newcommand{\hH}{\widehat{H}}

\begin{document}
\title{Towards Information Privacy for the Internet of Things}

\author{Meng~Sun,~\IEEEmembership{Student~Member,~IEEE},
        Wee~Peng~Tay,~\IEEEmembership{Senior~Member,~IEEE,
				and Xin~He}
\thanks{This research is supported in part by the Singapore Ministry of Education Academic Research Fund Tier 2 grant MOE2014-T2-1-028.}
\thanks{The authors are with the Department of Electrical and Electronic Engineering, Nanyang Technological University, Singapore, e-mails: MSUN002@e.ntu.edu.sg, wptay@ntu.edu.sg, hexin@ntu.edu.sg}
}

\maketitle
\begin{abstract}
In an Internet of Things network, multiple sensors send information to a fusion center for it to infer a public hypothesis of interest. However, the same sensor information may be used by the fusion center to make inferences of a private nature that the sensors wish to protect. To model this, we adopt a decentralized hypothesis testing framework with binary public and private hypotheses. Each sensor makes a private observation and utilizes a local sensor decision rule or privacy mapping to summarize that observation independently of the other sensors. The local decision made by a sensor is then sent to the fusion center. Without assuming knowledge of the joint distribution of the sensor observations and hypotheses, we adopt a nonparametric learning approach to design local privacy mappings. We introduce the concept of an empirical normalized risk, which provides a theoretical guarantee for the network to achieve information privacy for the private hypothesis with high probability when the number of training samples is large. We develop iterative optimization algorithms to determine an appropriate privacy threshold and the best sensor privacy mappings, and show that they converge. Finally, we extend our approach to the case of a private multiple hypothesis. Numerical results on both synthetic and real data sets suggest that our proposed approach yields low error rates for inferring the public hypothesis, but high error rates for detecting the private hypothesis.
\end{abstract}


\section{Introduction}
\label{sec:intro}

Sensor networks have seen widespread applications in industrial, military, and civilian monitoring applications like intrusion detection, target tracking, leakage detection and fall detection \cite{Butun2014,Chen2004,Stoianov2007,Chen2006,TayTsiWin:J08a,Alemdar2010,XuQuiLen:J15}. In the emerging Internet of Things (IoT) paradigm, large numbers of sensors are deployed to enable sense-making and intelligent analytics based on the sensors' observations. This can be modeled using the decentralized detection framework \cite{Tsi:93,VisVar:97,TayTsiWin:J08b,Tay:J12,CheCheVar:12,Tay:J15,HoTayQue:J15}, where each sensor makes an observation, summarizes this observation using a local decision rule, and sends the summary to a fusion center. Based on the received sensor summaries, the fusion center then makes the final inference on a phenomenon of interest.

While the fusion center's role is to perform inference on a particular hypothesis of interest, there is nothing stopping it from using the received sensor information to infer another correlated hypothesis. An example is the deployment of home-monitoring video cameras in old folks' homes for fall detection. If the cameras transmit the raw video feed to a fusion center, the fusion center can not only use these video feeds for fall detection, but also has the potential to intrude on the privacy of the home inhabitants. The camera sensors therefore need to perform intelligent observation summary with a suitable sensor rule in order to limit the amount and quality of information they send to the fusion center. Another example is when an insurance company wishes to determine if a person has a particular pre-existing medical condition using medical records from hospitals the person has been treated at. However, these medical records may reveal more than the particular condition that the insurance company is investigating. The hospitals will need to decide what to send to the insurance company to avoid disclosing the person's other medical conditions. Although this latter example is not in the context of sensor networks, we can see that it nevertheless falls in the framework of how to preserve privacy in decentralized detection while still enabling the fusion center to make an inference on a particular hypothesis. In this paper, we call a hypothesis a public hypothesis if its inference or detection is to be achieved by a sensor network specifically designed for this purpose. We call a hypothesis a private hypothesis if it can also be inferred based on the same sensor observations, but whose true state the sensor network wishes to protect. We call preventing the accurate inference of the private hypothesis \emph{information privacy} (see Section~\ref{sec:problem formulation} for a precise technical definition).

The main focus of this paper is to protect information privacy by making it difficult for the fusion center to perform inference on the private hypothesis. It has been shown in \cite{Tsi:88,ChaVee:03,Aldosari2004} that the error decay rate at the fusion center increases with the quality of information that the sensors convey to it. Therefore, by appropriately ``degrading'' the information sent by each sensor to the fusion center, we aim to achieve a good tradeoff between privacy leakage and the ability of the fusion center to infer the public hypothesis.

\subsection{Related Work}

%One encryption approach is homomorphic encryption \cite{Boneh2005,Ishai2007,Gentry2009}, which allows the manipulation of encrypted data without decrypting it. However, such techniques are restricted to a limited set of operations and have high complexities and communication requirements. Differential privacy \cite{Dwork2008differential} is a privacy guarantee that has been adopted by various works \cite{Friedman2010data,Andres2013geo,Chen2014correlated} to provide data privacy through addition of perturbation to sensor data. It allows the release of statistical information with rigorous privacy guarantees of individual sensor datum. Although both methods mentioned above provide data privacy against the fusion center, ensuring

With ubiquitous IoT devices monitoring every aspect of a user's life, privacy is a main consideration for users when adopting IoT technologies. Sensitive personal data like lifestyle preferences and location information may be abused for unwanted advertisement purposes or for more nefarious objectives like unauthorized surveillance. Privacy in IoT networks can be classified as data privacy and inference privacy.
%<*citations>
Data privacy refers to the protection of sensor data from unauthorized parties. Achieving data privacy has been comprehensively addressed by methods typically involving encryption or perturbation \cite{Boneh2005,Ishai2007,Gentry2009,Dwork2008differential,Friedman2010data,Andres2013geo,Chen2014correlated,duchi2013local}.
%</citations>
Although data privacy ensures that each sensor datum is protected, it does not stop a fusion center from inferring about a private hypothesis if statistical information about it is still present in the aggregated data.

%<*works1>
The focus of this paper is inference privacy, whose aim is to prevent the fusion center from using its received information to accurately infer a private hypothesis. The paper \cite{Yamamoto1983} utilized source coding so that a receiver can decode one source within a prescribed distortion tolerance, while ensuring that the mutual information between the decoded sequence and another correlated private source is lower than a threshold. The author also analyzed the privacy-utility tradeoff under this formulation. The reference \cite{sankar2010theory} extends the result of \cite{Yamamoto1983}, and designed a framework to quantify the privacy-utility tradeoffs. An information-theoretic scheme for information privacy was proposed by \cite{PinCalmon2012}, which formulated a convex program to find the privacy-preserving mapping to minimize the mutual information between the private hypothesis and information received at the fusion center, while satisfying certain utility constraints. In \cite{Li2014}, the minimum Bayesian error probability for a fusion center to infer a private phenomenon is used as the privacy metric, and a person-by-person optimization approach is proposed to find sensor decision rules, but inference privacy is not guaranteed in general.
%</works1>
All the aforementioned approaches are designed to achieve a good utility-privacy tradeoff, but assume knowledge of the joint distribution of sensor observations, public and private hypotheses. If a mismatched joint distribution is utilized, the utility-privacy tradeoff is impacted \cite{Salamatian2015}. Knowing the underlying joint distribution may not be practical in an IoT network. Therefore, in this paper, we propose a nonparametric hypothesis testing approach with information privacy constraints.

%<*works2>
A nonparametric approach to inference privacy was proposed by \cite{diamantaras2016data, al2017ratio}, in which sensor data is mapped to a subspace before being made available to the fusion center. These methods can be adopted in a centralized platform like cloud computing, but impractical for a sensor network because it requires the use of a \emph{trusted data curator} to first aggregate observations from all sensors, and then performing the privacy mapping on the aggregated data. Furthermore, no theoretical guarantees of the level of privacy achievable are provided in \cite{diamantaras2016data, al2017ratio}. In a decentralized architecture like an IoT network, each sensor makes its local decision based solely on its own observation, and independently of the other sensors.
%</works2>
A nonparametric decentralized detection method was introduced by \cite{NguWaiJor:05}, which proposed the use of a kernel-based method to learn the optimal sensor decision rules from a given set of labeled training data. Subsequently, \cite{Wang2016} extended this method using a weighted kernel to allow sensor selection in the decentralized detection procedure.  These works however do not address the inference privacy issue described above. Following these works, we assume that a set of labeled training data is available and employ a kernel-based approach to learn the sensor rules subject to a  privacy metric called information privacy \cite{PinCalmon2012}.

\subsection{Our Contributions}\label{subsec:contributions}

In this paper, we develop an information privacy preserving framework for nonparametric hypothesis testing and algorithms to realize our framework. Our main contributions are as follows:
\begin{enumerate}
		\item We adopt the concept of $\epsilon$-information privacy from \cite{PinCalmon2012}, and show that a sufficient condition to achieve $\epsilon$-information privacy when the private hypothesis is binary, is to ensure that the average of the Type I and II detection error probabilities are large. We show that contrary to intuition, a large Bayes detection error for the private hypothesis does not ensure information privacy.
		\item Since we do not assume knowledge of the underlying joint distribution and we adopt a nonparametric optimization framework, we introduce the concept of $(\epsilon,\delta)$-information privacy, which is a weak form of $\epsilon$-information privacy. When the private hypothesis is binary, we propose a privacy metric constraint, which we call the \emph{empirical normalized risk}, and show that under some mild technical assumptions, this achieves $(\epsilon,\delta)$-information privacy for any $\delta>0$ when the training sample size becomes large.
    \item When both public and private hypotheses are binary, we propose a nonparametric privacy-aware optimization framework and iterative algorithms to learn an appropriate privacy threshold for our empirical normalized risk, and the optimal sensor privacy mappings. We show that both algorithms converge to the critical points of their respective objective functions.
		\item We extend our optimization framework to the case where the private hypothesis is $m$-ary with $m> 2$, and provide a sufficient condition to achieve information privacy. This condition translates into $m-1$ empirical normalized risk constraints in our optimization framework, which achieves $(\epsilon,\delta)$-information privacy for any $\delta>0$ when the training sample size becomes large.
    \item We verify the performance of our algorithm on both simulated data and real data. Our experiments suggest that our approach can achieve low error rates for inferring the public hypothesis, but high error rates for detecting the private hypothesis.
\end{enumerate}

This paper is an extension of our conference paper \cite{SunTay:C16}, which utilized a nonparametric approach to learn sensor decision rules under a Bayesian error probability privacy constraint. As shown in this paper, that approach does not guarantee information privacy in general.

The rest of this paper is organized as follows. In Section~\ref{sec:problem formulation}, we present our system model and assumptions. We also define information privacy and present a sufficient condition to achieve it. In Section~\ref{sec:algorithm}, we propose a nonparametric privacy-aware optimization framework based on an empirical normalized risk constraint, to obtain the sensor privacy mappings. We show that our approach achieves information privacy with high probability as the training sample size becomes large. We then propose an iterative algorithm for the optimization problem. We present simulation results to verify the effectiveness of the proposed algorithms in Section~\ref{sec:simulation}, and we conclude in Section~\ref{sec:conclusions}.

\emph{Notations:}
We use capital letters like $X$ to denote random variables or vectors, lowercase letters like $x$ for deterministic scalers, and boldface lowercase letters like $\bx$ for deterministic vectors. We use $\Real$ to denote the set of real numbers, and $\Gamma^c$ to be the complement of the set $\Gamma$. The indicator function $\indicator{A}$ takes value 1 iff the clause $A$ is true. We let $(x)_+=\max\{x,0\}$. We assume that all random variables are defined on the same underlying probability measure space with probability measure $\P$, and $\E$ is the associated expectation operator. In some cases, for clarity, we use $\E_X$ to emphasize that the expectation is with respect to (w.r.t.) $X$. We use $p_X(\cdot)$ to denote the probability mass function of $X$, and $p_{X\mid Y}(\cdot \mid \cdot)$ to denote the conditional probability mass function of $X$ given $Y$.

\section{Problem Formulation and Information Privacy}
\label{sec:problem formulation}
In this section, we first describe our system model and assumptions, and then discuss the connection of our problem setup with information privacy and differential privacy. We provide a sufficient condition to achieve information privacy for an IoT network.

\subsection{System Model}\label{subsec:system}
\begin{figure}[!htb]
\centering
\includegraphics[width=0.45\textwidth]{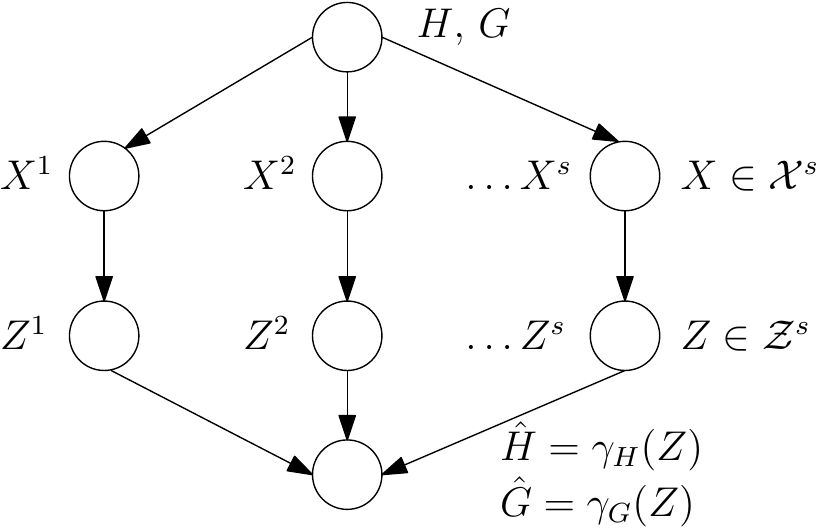}
\caption{An IoT network with public hypothesis $H$ and private hypothesis $G$.}
\label{fig:PBL}
\end{figure}

We consider a decentralized detection network as shown in Fig.\ref{fig:PBL}. Suppose that two hypotheses $H$ and $G$ each takes binary values in $\{-1,+1\}$. Each sensor $t\in\{1,2,\ldots,s\}$, makes a noisy observation $X^t \in \calX$ of $(H,G)$, where $\calX =\{1,2,\ldots,|\calX|\}$. It then summarizes its observation using a local decision rule or privacy mapping $\gamma^{t}: \calX \mapsto \calZ$, with $\calZ=\{1,2,\ldots,|\calZ|\}$, and transmits $Z^t = \gamma^t(X^t)$ to a fusion center. Each sensor's local decision rule is allowed to be a probabilistic mapping from $\calX$ to $\calZ$. The transmission from each sensor to the fusion center is constrained to a limited number of bits, so that $|\calX| \geq |\calZ|$. This models an ad hoc IoT network with low-power devices that may be battery operated. For example, the NB-IoT standard \cite{NB-IOT_Huawei} is developed for low-cost and low-power devices with limited communication bandwidths. In this paper, we do not require that the sensor observations $X^t$, $t=1,\ldots,s$ are independent.

Let $X =(X^{1},\ldots,X^{s})$ and $Z=(Z^{1},\ldots,Z^{s})$. Based on the received messages $Z$, the fusion center makes a decision $\hH=\gamma_H(Z) \in \{-1,+1\}$ about the state of the hypothesis $H$. We consider $H$ to be the public hypothesis that the sensors want the fusion center to infer correctly. On the other hand, $G$ is a private hypothesis that the sensors wish to hide from the fusion center. The fusion center however is curious, and after receiving the local decisions from the sensors, implements a decision rule $\hG=\gamma_G(Z) \in \{-1,+1\}$ to infer the private hypothesis $G$. Our goal is to find, for each sensor $t=1,\ldots,s$, a local decision rule or privacy mapping $\gamma^{t}$ to minimize the error probability $\P(H\neq \gamma_H(Z))$, while making it difficult for any fusion rule $\gamma_G$ the fusion center may employ to detect $G$.

In the following Section~\ref{subsec:information_privacy}, we show that by making the average of the Type I error $\P(\gamma_G(Z) = -1 \mid G=1)$ and Type II error $\P(\gamma_G(Z) = 1 \mid G=-1)$ sufficiently large, we achieve \emph{information privacy} \cite{PinCalmon2012} for the private hypothesis $G$, which implies \emph{differential privacy} \cite{Dwork2008differential}. However, to find the optimal local sensor decision rules in order to facilitate the inference of $H$ while ensuring that both the Type I and II errors are  large, one needs to know the joint probability distribution $p_{X,H,G}$ a priori. This assumption is impractical for many IoT applications as the underlying joint distribution may be difficult to model accurately. Therefore, we assume that this distribution is unknown, but we are given a set of $n$ independent and identically distributed (i.i.d.) training data $(\bx_{i},h_{i},g_{i})_{i=1}^{n}$ sampled from $p_{X,H,G}$.\footnote{In the sequel, we adopt the following convention: a superscript index corresponds to a sensor index (e.g., $X^t$), while a subscript index corresponds to either a training data index (e.g., $\bx_i$) or a hypothesis.}  We adopt the framework of empirical risk minimization as in \cite{NguWaiJor:05} to design the local sensor decision rules, which we discuss in detail in Section~\ref{subsec:nonparametric framework}.

\subsection{Information Privacy}\label{subsec:information_privacy}

To prevent the fusion center from inferring the true state of the private hypothesis $G$ based on the received sensor messages $Z$, we need to design $Z$ so that the prior and posterior distributions of $G$ are close to each other. This is the definition of $\epsilon$-information privacy given by \cite{PinCalmon2012}, which we recall below. Note that this differs from the more commonly used differential privacy concept \cite{Dwork2008differential,Shi2011privacy}, which \cite{PinCalmon2012} shows is weaker than information privacy in general.

\begin{Definition}[Information privacy]\label{def:info privacy}
Let $G$ and $Z$ be random variables. For $\epsilon > 0$, $G$ given $Z$ or $p_{G\mid Z}$ has $\epsilon$-information privacy, if for almost surely any $(G,Z)=(g, \bz)$, we have
\begin{align}\label{ineq:info privacy}
    e^{-\epsilon}\leq\frac{p_{G \mid Z}( g \mid \bz)}{p_{G}(g)}\leq e^{\epsilon}.
\end{align}
The value $\epsilon$ is called the information privacy budget.
\end{Definition}

In an IoT network, both data and information privacy are important design aspects. Achieving data privacy has been comprehensively addressed in \cite{Ziegeldorf2014privacy,duchi2013local,chaudhuri2011differentially,duchi2014privacy,hall2013differential}. In this paper, we focus on the information privacy aspect. In \cite{PinCalmon2012}, an optimization framework based on minimizing the mutual information between $G$ and $Z$ is proposed. This leads to a weaker privacy guarantee than information privacy. In the following, we show that if $G$ is a binary hypothesis, then constraining the total probability of error in detecting $G$ leads to information privacy for $G$ under some technical conditions. We make following assumption.

\begin{Assumption}\label{assumpt:regular}
The priors $p_G(1), p_G(-1) > 0$. The supports of the conditional distributions of $Z$ given $G=1$ and $G=-1$ are the same, and denoted as $\calD$.
\end{Assumption}

Assumption~\ref{assumpt:regular} leads to no loss of generality and is required to avoid trivial cases where $G$ is perfectly detectable. For any detector $\gamma(Z)$ for $G$ based on $Z$, let
\begin{align*}
R(\gamma) &= \ofrac{2}(\P(\gamma(Z)=-1\mid G=1)+\P(\gamma(Z)=1\mid G=-1))
\end{align*}
be the average of the Type I and II error probabilities. For each $\bz \in \calD$, let
\begin{align*}
\ell(\bz) = \frac{p_{Z\mid G}(\bz \mid 1)}{p_{Z\mid G}(\bz \mid -1)},
\end{align*}
and
\begin{align}
c = \min\Big\{& \P(\ell(\bz)=\min_{\bz\in\calD}\ell(\bz)){G=-1},\nonumber\\
& \quad\P(\ell(\bz)=\max_{\bz\in\calD}\ell(\bz)){G=1}\Big\}.\label{eqn:c}
\end{align}
From Assumption~\ref{assumpt:regular}, since $0< \ell(\bz) < \infty$ for all $\bz\in\calD$, $c > 0$.

The following result shows that a small information privacy budget is essentially equivalent to a large $\min_\gamma R(\gamma)$. Since using (an empirical proxy of) $\min_\gamma R(\gamma)$ as a privacy constraint is more convenient than imposing \eqref{ineq:info privacy} directly (which leads to $2|\calZ|$ constraints when $G$ is binary), the following proposition allows us to formulate a compact privacy constraint in Section~\ref{sec:algorithm}.

\begin{Proposition}\label{thm:PrivacyMetrics}
Suppose that Assumption~\ref{assumpt:regular} holds.
\begin{enumerate}[(i)]
    \item \label{it:bayes-infoprivacy} Let $\epsilon >0$ be a sufficiently small positive constant. If $G$ given $Z$ has $\epsilon$-information privacy, then $R(\gamma_G)\geq \theta/(2\max_g p_G(g))$, for all $\theta \in [0,1/2]$ that satisfies
			\begin{align}\label{ineq:delta-infoprivacy}
			\calH(\theta)\leq \calH(G)-\epsilon,
			\end{align}
			with $\calH(\cdot)$ being the binary entropy function \cite{Cover2006}.

    \item \label{it:infoprivacy-bayes}
    If $\min_{\gamma}R(\gamma) \geq \theta\in[0,1/2]$, $G$ given $Z$ achieves $\epsilon$-information privacy where $\epsilon = \log \frac{c}{(c+2\theta-1)_+}$.
\end{enumerate}
\end{Proposition}
\begin{IEEEproof}
See Appendix~\ref{prf:PrivacyMetrics}.
\end{IEEEproof}

If the prior probabilities $p_G(1)=p_G(-1)=1/2$, $\min_\gamma R(\gamma)$ is the Bayesian error probability. When the prior for $G$ is not uniform, the Bayes detector $\gamma_G^*$ with a large Bayes error $\P(\gamma_G^*(Z)\ne G)$ do not guarantee information privacy as defined in Definition~\ref{def:info privacy}. An example is shown in Example~\ref{eg:Bayes vs InfoPrivacy}, and a simulation to demonstrate this is given in Section~\ref{subsubsec:compare}. To ensure information privacy, we require that both the Type I and II error probabilities are large, which is equivalent to having the risk $\min_\gamma R(\gamma)$ to be sufficiently large.

\begin{Example}\label{eg:Bayes vs InfoPrivacy}
Suppose $p_G(-1)<1/2$, and $G$ and $Z \in \{\bz_1,\bz_2\}$ have the joint distribution shown in Table~\ref{tab:distribution}. Then, for all sufficiently large $d$, the optimal Bayesian error probability detector declares $G^*=1$ for all $Z$. The Bayes error $\P(G^*\neq G)=p_G(-1)$, which is the largest possible Bayes error for any distribution $p_{Z\mid G}$. However, we have $\frac{p_{G\mid Z}(-1\mid \bz_1)}{p_{G}(-1)}=\ofrac{p_G(-1)+ d(1-2p_{G}(-1))}\to 0$ as $d\to\infty$, which invalidates \eqref{ineq:info privacy} for any $\epsilon \in [0,\infty)$. This shows that a large Bayesian error probability does not guarantee information privacy of $G$.

%In this example, we also have the mutual information $\calI(G;Z) \to 0$ as $d\to\infty$ and $p_G(-1)\to 0$. It shows that for any $\delta > 0$ and $\epsilon \in [0,\infty)$, one can find a joint distribution $p_{G,Z}$ such that $\calI(G;Z)\leq\delta$ while \eqref{ineq:info privacy} is not satisfied. Therefore, a small $\calI(G;Z)$ (also known as mutual information privacy \red{[citation]}) does not guarantee information privacy in general.\red{[Not valid example.]}
\end{Example}

%If we use the maximum likelihood detector, we have $R(\gamma_G)=\ofrac{2}(\ofrac{d}+\frac{p_{G}(-1)}{p_{G}(1)})\to 0$. This is in accord with our theorem, since From Theorem~\ref{thm:PrivacyMetrics}\ref{it:infoprivacy-bayes}, we know that if $R(\gamma_G)$ is large, information privacy will be protected.

\begin{table}[!ht]
\centering
\caption{Joint distribution of $Z$ and $G$}
\label{tab:distribution}
\begin{tabular}{|c|c|c|c|}
\hline
\multicolumn{2}{|c|}{\multirow{2}{*}{$P_{G,Z}(g,\bz)$}} & \multicolumn{2}{c|}{$Z$}        \\ \cline{3-4}
\multicolumn{2}{|c|}{}            & $\bz_1$     & $\bz_2$         \\ \hline
\multirow{2}{*}{$G$}              & $-1$ & $p_G(-1)/d$   & $(1-1/d)p_G(-1)$     \\ \cline{2-4}
                                  & $1$                    & $p_G(1)-p_G(-1)$ & $p_G(-1)$ \\ \hline
\end{tabular}
\end{table}

\section{Privacy Mapping Design}
\label{sec:algorithm}

In this section, we formulate a nonparametric privacy-aware optimization framework to find privacy mappings or local decision rules for the sensors in order to achieve information privacy for the private hypothesis $G$. We show that our approach achieves a weak form of information privacy, and we propose iterative algorithms to solve the optimization problem.

\subsection{Nonparametric Privacy-aware Optimization}
\label{subsec:nonparametric framework}

From Proposition~\ref{thm:PrivacyMetrics}\ref{it:infoprivacy-bayes}, we can solve the following optimization problem to obtain sensor privacy mappings $\{\gamma^t : t=1,\ldots,s\}$ in order to achieve a desired level of information privacy for $G$:
\begin{align}\label{gop}
\begin{aligned}
\min_{\gamma^1,\ldots,\gamma^s,\gamma_H}& \P(\gamma_H(Z)\neq H)\\
\text{s.t. }\ & \min_{\gamma_G}R(\gamma_G) \geq \theta,
\end{aligned}
\end{align}
where $\theta$ is a privacy threshold that determines the information privacy budget for $G$.
However, since \eqref{gop} requires knowledge of the joint distribution of sensor observations and the hypotheses $H$ and $G$, such an approach may not be practical in IoT applications. Instead, we propose a nonparametric approach in which we use empirical regularized risks to approximate the Bayesian error probability for $H$ and the risk $R(\gamma_G)$ for $G$.

Let $\scH$ be a reproducing kernel Hilbert space (RKHS) associated with a kernel $\kappa(\cdot,\cdot)$. Let $\ip{\cdot}{\cdot}$ be the kernel inner product associated with $\kappa(\cdot,\cdot)$, and $\norm{\cdot}$ denote the norm induced by the kernel inner product. Following \cite{NguWaiJor:05}, we restrict $\gamma_H$ and $\gamma_G$ to be from $\scH$, i.e., $\gamma_H$ and $\gamma_G$ are of the form
\begin{align*}
    \gamma_H(\bz) &= \ip{w_H(\cdot)}{\Phi(\bz)(\cdot)},\\
    \gamma_G(\bz) &= \ip{w_G(\cdot)}{\Phi(\bz)(\cdot)},
\end{align*}
where $w_H(\cdot), w_G(\cdot) \in \scH$, and $\Phi(\bz)(\cdot)=\kappa(\cdot,\bz)$ is the feature map, which maps a point $\bz$ to a function $\Phi(\bz)(\cdot):\calZ^s\mapsto\Real$. To simplify notations, we use $w_H,w_G$, and $\Phi(\bz)$ respectively in this paper.

For each $t=1,\ldots,s$, $x^{t}\in\calX$, and $z^{t}\in\calZ$, let $Q^{t}(z^{t} \mid x^{t}) = \P(Z^t = z^t){X^t = x^t}$  represent the privacy mapping $\gamma^t$. Let $Q(\bz\mid \bx) = \P(Z = \bz \mid X = \bx)$ denote the overall sensor mapping, and $\calQ$ be the set of $Q(\cdot \mid \cdot)$ such that
%<*Q>
\begin{align}
    &Q(\bz\mid \bx) = \prod_{t=1}^{s} Q^{t}(z^{t} \mid x^{t}),\label{Q}\\
    &\sum_{z^t\in\calZ} Q^{t}(z^t \mid  x^t) =1,\label{intQ}\\
    &Q^{t}(z^{t} \mid x^{t})\geq 0.\label{Qnonneg}
\end{align}
%</Q>

Let $\phi$ be a loss function. We seek to minimize the regularized empirical $\phi$-risk of deciding $H$, while ensuring that the \emph{empirical normalized $\phi$-risk} of deciding $G$, is higher than a given threshold $\theta$. We assume that we are given $n$ i.i.d.\ training data points $(x_{i},h_{i},g_{i})_{i=1}^{n}$, and we seek to
\begin{subequations}\label{original_opt}
\begin{align}\label{minH1}
	\begin{aligned}
		\min_{w_H\in\scH,Q\in\calQ}\ofrac{n}\sum_{i=1}^{n} \sum_{\bz\in\calZ^s} & \phi(h_{i}\ip{w_H}{\Phi(\bz)})Q(\bz \mid \bx_{i})\\
		&+\frac{\lambda}{2}\norm{w_H}^2,
	\end{aligned}
\end{align}
    \begin{align}\label{constG1}
			\begin{aligned}
				\text{s.t. } \min_{w_G\in\scH}\ofrac{2}\sum_{g\in\{-1,1\}}\sum_{i\in\calS_{g,n}}\sum_{\bz\in\calZ^s} & \frac{\phi(g_i\ip{w_G}{\Phi(\bz)})Q(\bz \mid \bx_{i})}{|\calS_{g,n}|} \\
				&+\frac{\lambda_n}{2}\norm{w_G}^2\geq \theta.
			\end{aligned}
		\end{align}
\end{subequations}
Here, $\calS_{g,n}=\{i\in\{1,\ldots,n\}: g_i=g\}$, for $g\in\{-1,1\}$, and $\lambda >0$ and $\lambda_n>0$ are regularization weights. The regularized empirical risks are used, since without the regularization, it is known that the generalization error can become large \cite{Scholkopf2001}. Note also that the empirical normalized risk in \eqref{constG1} is different from the traditional empirical risk used in the machine learning literature (cf.\ \eqref{minH1} and \cite{Scholkopf2001}), and serves as a surrogate for $R(\cdot)$ in Proposition~\ref{thm:PrivacyMetrics}.

The optimization problem \eqref{original_opt} is intractable to compute because of the summation over $\calZ^s$, therefore we employ the same lower bound relaxation as in (15) of \cite{NguWaiJor:05}, and let $\Phi_Q(\bx)=\sum_{\bz\in\calZ^s}Q(\bz|\bx)\Phi(\bz)$. The corresponding kernel is  $\kappa_{Q}(\bx,\bx')=\ip{\Phi_Q(\bx)}{\Phi_Q(\bx')}=\sum_{\bz}\sum_{\bz'} Q(\bz\mid\bx)Q(\bz'\mid\bx')\kappa(\bz,\bz')$. With this relaxation, the convex surrogate of the risk $R(\gamma_G)$ becomes
\begin{align}\label{Rphi}
\begin{aligned}
    R_\phi(w_G,Q)=&\ofrac{2}\E{X}[\phi(-\ip{w_G}{\Phi_Q(X)})\mid G=-1]\\
    &+\ofrac{2}\E{X}[\phi(\ip{w_G}{\Phi_Q(X)})\mid G=1],
\end{aligned}
\end{align}
and the surrogates for the empirical risks for $H$ and $G$ are respectively,
\begin{align}
    F(w_H,Q)=\ofrac{n}\sum_{i=1}^n\phi\left(h_{i}\ip{w_H}{\Phi_Q(\bx_i)}\right)+\frac{\lambda}{2}\norm{w_H}^{2},\label{FH}\\
    \begin{aligned}\label{FG}
    \hat{R}_n(w_G,Q)=&\ofrac{2}\sum_{g\in\{-1,1\}}\sum_{i\in\calS_{g,n}}\frac{\phi(g_i\ip{w_G}{\Phi_Q(\bx_i)})}{|\calS_{g,n}|}\\
    &+\frac{\lambda_n}{2}\norm{w_G}^2.
    \end{aligned}
\end{align}
Our optimization problem \eqref{original_opt} is then relaxed to
%<*final>
\begin{subequations}\label{final0}
\begin{align}
    \min_{w_H\in\scH, Q\in\calQ}\ & F(w_H,  Q)\label{finalF}\\
    \text{s.t. } & \min_{w_G\in\scH}\hat{R}_n(w_{G},Q)\geq\theta. \label{finalR}
\end{align}
\end{subequations}
%</final>
For convenience, we call \eqref{final0} the nonparametric privacy-aware optimization (NPO) framework. We note that \eqref{constG1} holds if the constraint \eqref{finalR} holds, since \eqref{finalR} is a lower bound relaxation. In general, \eqref{finalR} cannot guarantee information privacy with probability one since \eqref{final0} is solved using a finite number of training data points $(x_{i},h_{i},g_{i})_{i=1}^{n}$. Therefore, we introduce the notion of \emph{weak} information privacy below.

\begin{Definition}[$(\epsilon,\delta)$-Information privacy]\label{def:loose info privacy}
Let $G$ and $Z$ be random variables. For $\epsilon > 0$, $\delta \in [0,1]$, $G$ given $Z$ or $p_{G\mid Z}$ has $(\epsilon,\delta)$-information privacy, if
\begin{align*}%\label{ineq:loose info privacy}
\P(e^{-\epsilon}\leq\frac{p_{G \mid Z}( G \mid Z)}{p_{G}(G)}\leq e^{\epsilon})\geq 1-\delta.
\end{align*}
\end{Definition}

Our goal is to provide a theoretical guarantee for weak information privacy with some $\delta\in(0,1)$ using \eqref{finalR}. However, since we are using a loss function $\phi$ as a surrogate for the 0-1 loss implied in $R(\cdot)$, we require the following assumptions, which have been used to bound generalization errors in \cite{Zhang2004}.

\begin{Assumption}[Convex loss function]\label{assumpt:loss}
The loss function $\phi$ is a convex function and $\phi(0) < \infty$.
\end{Assumption}

\begin{Assumption}\label{assumpt:bound}
For any $\eta\in[0,1]$, let $R_\phi^*(\eta)=\inf_{\gamma\in\Real}\left(\eta\phi(\gamma)+(1-\eta)\phi(-\gamma)\right)$. There exist $a>0$ and $r\geq 1$, such that for all $\eta\in[0,1]$,
\begin{align}\label{ineq:bound}
    a^r(\phi(0)-R_\phi^*(\eta))\geq \left|\ofrac{2}-\eta\right|^r.
\end{align}
\end{Assumption}

\begin{table*}[!t]
\centering
\caption{Loss functions, their conjugate duals, and parameters for Assumption~\ref{assumpt:bound}}
\label{tab:loss function}
\begin{tabular}{|c|c|c|c|c|}
    \hline
    Loss function $\phi$ & Conjugate dual $\phi^*(-x^*)$&Assumption~\ref{assumpt:bound}\\ \hline%&Decision rule of $R_\phi^*$
    $\phi(u)=e^{-u}$ & $\left\{\begin{array}{ll} x^*\ln x^* -x^*,&x^*\in(0,\infty)\\0,&x^*=0\end{array}\right.$&$a=\ofrac{\sqrt{2}}, r=2$\\ \hline%&$\ofrac{2}\ln \frac{\eta(\bz)}{1-\eta(\bz)}$
    $\phi(u)=\log(1+e^{-u})$            & $\left\{\begin{array}{ll}x^*\log x^*+(1-x^*)\log(1-x^*),&x^*\in(0,1)\\0,&x^*=0,1\end{array}\right.$&$a=\ofrac{\sqrt{2}}, r=2$\\ \hline%&$\ln \frac{\eta(\bz)}{1-\eta(\bz)}$
    $\phi(u)=\max\{1-u,0\}$             & $-x^* ,x^*\in[0,1]$                   &$a=1/2, r=1$\\ \hline%&$\sign (2\eta(\bz)-1)$
    $\phi(u)=(1-u)^2$                   & $(x^*)^2/4-x^*,x^*\in\Real$        &$a=1/2, r=2$\\ \hline%&$2\eta(\bz)-1$
\end{tabular}
\end{table*}

Examples of loss functions, with their corresponding $a$ and $r$ values in Assumption~\ref{assumpt:bound}, are shown in Table~\ref{tab:loss function}. These include commonly used loss functions like the logistic loss function \cite{Friedman2000}, exponential loss function \cite{Freund1995}, hinge loss function \cite{Rosasco2004loss}, and quadratic loss function, which have been shown to be computationally efficient and have bounded approximation and estimation errors \cite{Bartlett2006,Zhang2004}. In Table~\ref{tab:loss function} we also list the conjugate dual $\phi^*$ of these loss functions, which will be useful in the sequel.

For $\bz\in\calZ^s$, let $\eta(\bz) = \P(\widetilde{G}=1){Z=\bz}$, where $\widetilde{G}$ induces the same conditional distribution for $Z$ as $G$ but has uniform prior. It can be shown that $\min_\gamma R(\gamma) = 1/2 - \E[|1/2 - \eta(Z)|]$ (cf.\ \eqref{Rstar}). Therefore, Assumption \ref{assumpt:bound} gives a lower bound for $\min_\gamma R(\gamma)$ in terms of $\E[R_\phi^*(\eta(Z))]$. To relate this to the left hand side of \eqref{finalR}, we need the following assumption.

%\begin{Rem}
%We need Assumption~\ref{assumpt:bound} to related the Bayes error and the minimum $\phi$ risk $R_\phi^*(\eta(\bz))$. From Theorem~2.2 of \cite{Zhang2004}, we see in Assumption~\ref{assumpt:bound}, if $R_\phi^*(\eta(\bz))$ is differentiable, the left hand side of \eqref{ineq:bound} becomes $d_{R_\phi^*}(1/2,\eta(\bz))$, where $d_{R_\phi^*}$ is the distance specified by the Bergman divergence of $R_\phi^*$. The right hand side is the absolute difference between $1/2$ and $\eta(\bz)$. So this assumption intuitively holds, since both sides in \eqref{ineq:bound} change simultaneously.
%\end{Rem}

\begin{Assumption}\label{assumpt:phi2bayes}\
\begin{enumerate}[(i)]
    \item \label{it:assumpt}$\lambda_n\to 0$, as $n\to\infty$.
    \item \label{it:strictly positive} The kernel $\kappa(\cdot,\cdot)$ is a strictly positive kernel.
\end{enumerate}
\end{Assumption}

Examples of strictly positive kernels include the Gaussian kernel, and kernels defined by inverse multiquadrics \cite{wendland2004scattered}. With the above assumptions, the following result gives a theoretical guarantee of weak information privacy using the constraint \eqref{finalR}.

\begin{Theorem}\label{thm:empricalInfoPrivacy}
Suppose that Assumptions~\ref{assumpt:regular}--\ref{assumpt:phi2bayes} hold. Then for any $\delta\in(0,1]$, there exists $n_0$, such that for all $n\geq n_0$, $G$ achieves $(\epsilon,\delta)$-information privacy for any $Q\in\calQ$, if
\begin{align*}
\min_{w_G\in\scH}\hat{R}_n(\hat{w}_{G},Q)\geq\theta,
 \end{align*}
 where $\epsilon=\log\frac{c}{\left(c-2a(\phi(0)-\theta+\delta)^{1/r}\right)_+}$, with $c$ as defined in \eqref{eqn:c}.
\end{Theorem}
\begin{IEEEproof}
See Appendix~\ref{prf:empricalInfoPrivacy}.
\end{IEEEproof}

From Theorem~\ref{thm:empricalInfoPrivacy}, we see that our NPO \eqref{final0} can achieve arbitrarily strong information privacy with $\epsilon\to 0$ and $\delta\to 0$ by taking the sample size $n\to\infty$ and $\theta\to\phi(0)$. The latter condition will however result in a poor detection rate for $H$, since it leads to the case where all sensor observations are mapped to the same output. Therefore, in the following, we propose an iterative procedure to find a suitable threshold $\theta$ with appropriate constraints on the privacy mappings.

\subsection{Iterative Optimization}

Let $\bbeta=[\beta_1,\beta_2,\ldots,\beta_n]^T \in \Real^n$, and
\begin{align}\label{dualR}
\begin{aligned}
    \hat{R}^*(\bbeta,Q)=&-\sum_{g\in\{-1,1\}}\sum_{i\in\calS_{g,n}}\frac{\phi^{*}(-2|\calS_{g,n}|\beta_{i})}{2|\calS_{g,n}|}  \\     &-\ofrac{2\lambda_n}\sum_{i=1}^n\sum_{j=1}^n\beta_{i}\beta_{j}g_{i}g_{j}\kappa_{Q}(\bx_i,\bx_j),
\end{aligned}
\end{align}
where $\phi^{*}$ is the conjugate dual of $\phi$ (see Table~\ref{tab:loss function} for some examples) \cite{Rockafellar2015}. From Proposition 2 in \cite{NguWaiJor:05}, for any $Q$, we have
\begin{align}\label{primal-dual}
\min_{w_G\in\scH}\hat{R}_n(w_G,Q)=&\max_{\bbeta\in\Real^n}\hat{R}^*(\bbeta,Q).
\end{align}
The optimal primal variable $w_G$ and the optimal dual variable $\bbeta$ in \eqref{primal-dual} are related by $w_G=\sum_{i=1}^n\beta_{i}g_{i}\Phi_Q(\bx_{i})$. The constraint \eqref{finalR} can now be rewritten as $\max_{\bbeta\in\Real^n}\hat{R}^*(\bbeta, Q)\geq \theta$. By using the interior-point method with log barrier \cite{Boyd2004}, we transform the optimization problem \eqref{final0} into the following:
\begin{align}
    \min_{w_H\in\scH,\bbeta\in\Real^n, Q\in\calQ} F_{0}(w_H,\bbeta, Q),\label{F0}
\end{align}
where $F_{0}(w_H,\bbeta, Q) = F(w_H, Q) -\frac{1}{\mu}\log(\hat{R}^*(\bbeta, Q)-\theta)$ and $\mu > 0$ is the barrier parameter.

We propose a two-step algorithm to solve \eqref{F0}. Since its unclear how to choose the threshold $\theta$ a priori, in the first step, we use an iterative algorithm to find $\theta^{*}$, which is the maximal possible $\theta$, subject to some constraints on $Q$. The first step also provides the initial point for the iterative algorithm in the second step. In the second step, we set the threshold $\theta$ to be a fixed fraction $p$ of $\theta^*$, and iteratively solve \eqref{F0}. We call $p$ the privacy threshold ratio. In both steps, we apply a \emph{block Gauss-Seidel method} \cite{Grippo2000}. We describe our algorithm in detail as follows.

\subsubsection{Finding the privacy threshold}

In Algorithm~\ref{algo:step1}, we apply the block Gauss-Seidel method to solve
\begin{align}
\max_{\bbeta\in\Real^n,Q\in\calQ'}\hat{R}^*(\bbeta,Q),\label{step1}
\end{align}
where $\calQ'$ is the set of $Q\in\calQ$, such that for all $t=1,\ldots,s$,
\begin{align}
    &\sum_{x\in\calX} Q^{t}(z \mid x)\geq \Delta_1,\ \text{for all $z\in\calZ$}, \label{Qnsmall}\\
    &\left|Q^{t}(z \mid x)-\ofrac{|\calZ|}\right|\geq \Delta_2,\ \text{for all $z\in\calZ$ and $x\in\calX$,} \label{Qnequal}
\end{align}
with $\Delta_1$ and $\Delta_2$ being small positive constants. The constraint \eqref{Qnsmall} ensures that no $z\in\calZ$ has small probability (otherwise we could have reduced the size of $\calZ$), and \eqref{Qnequal} prevents equal probabilities from being assigned to all possible $z$. Since the optimization problem in \eqref{step1} is non-convex, there is no guarantee that the block Gauss-Seidel method converges to the global optimum \cite{Xu2013,Tseng2001,Grippo2000}. However, we can show the following convergence result.
\begin{Proposition}
\label{pro:converge1}
Algorithm~\ref{algo:step1} converges a critical point.
\end{Proposition}
\begin{IEEEproof}
See Appendix~\ref{prf:converge1}
\end{IEEEproof}

The objective value output from Algorithm~\ref{algo:step1} is denoted as $\theta^*$, and we use its solution $(\widetilde{\bbeta},\widetilde{Q})$ as the initialization point in the second step.

\begin{algorithm}
\caption{Finding $\theta^{*}$}
\label{algo:step1}
    \begin{algorithmic}[1]
    \STATE{\textbf{input:} $\{g_{i},x^{1}_{i},\ldots,x^{s}_{i}\}_{i=1}^{n}$}
    \STATE{\textbf{initialization:} $\bbeta[0]\in\Real^n, Q[0]\in\calQ'$, $k=0$,}
    \REPEAT
    \STATE{
        \begin{itemize}[leftmargin=*]
            \item $k=k+1$,
            \item fix $ Q[k-1]$, solve the following convex optimization problem,
                \begin{align*}
                    \bbeta[k]=\argmax_{\bbeta\in\Real^n}\hat{R}^*(\bbeta, Q[k-1]),
                \end{align*}
            \item fix $\bbeta[k]$, $ Q^{j}[k]$, for $j<t$, and $ Q^j[k-1]$, for $j>t$, $t=1,2,\ldots,s$ update
                \begin{align*}
                     Q^{t}[k] =&\argmax_{ Q^{t}\in\calQ'}\hat{R}^*(\bbeta[k], Q^{1}[k],\ldots Q^{t-1}[k], \\
                     &Q^{t}, Q^{t+1}[k-1],\ldots, Q^{s}[k-1]),
                \end{align*}
        \end{itemize}
    }

    \UNTIL{
        $\frac{\hat{R}^*(\bbeta[k], Q[k])-\hat{R}^*(\bbeta[k-1], Q[k-1])}{\hat{R}^*(\bbeta[k-1], Q[k-1])}\leq\epsilon$,
    }
    \RETURN $\widetilde{\bbeta}=\bbeta[k],\widetilde{Q}= Q[k],\theta^*=\hat{R}^*(\bbeta[k], Q[k])$.

    \end{algorithmic}
\end{algorithm}

\subsubsection{Finding sensor privacy mappings}

With $\theta^*$ from Algorithm~\ref{algo:step1}, we set $\theta=p\theta^{*}$, where $p\in(0,1)$ is a constant typically chosen to be close to 1 (see Section~\ref{subsec:synthetic}). We minimize $F_{0}$ over $(w_H,\bbeta, Q)$ using a \emph{block Gauss-Seidel method}, as shown in Algorithm~\ref{algo:step2}.

Let $\balpha=[\alpha_1,\alpha_2,\ldots,\alpha_n]^T \in \Real^n$, and
\begin{align*}
    F^*(\balpha,Q)=&-\ofrac{n}\sum_{i=1}^n\phi^{*}(-n\alpha_{i})\\
    & -\ofrac{2\lambda}\sum_{i=1}^n\sum_{j=1}^n\alpha_{i}\alpha_{j}h_{i}h_{j}\kappa_{Q}(\bx_i,\bx_j).
\end{align*}
We have
\begin{align}
\min_{w_H\in\scH}F(w_H,Q)=&\max_{\balpha\in\Real^n}F^*(\balpha,Q),\label{dualalpha}
\end{align}
where the optimal primal variable $w_H$ and the optimal dual variable $\balpha$ in \eqref{dualalpha} are related by

\begin{align}\label{eq:optimal_w_alpha}
w_H=\sum_{i=1}^n\alpha_{i}h_{i}\Phi_Q(\bx_{i}).
\end{align}

Instead of recording the optimal $w_H$, which is associated with the feature map, we record the value of $\balpha$, which is a vector of length $n$. Note that if we substitute the updated $w_H$ with $\balpha$ according to \eqref{eq:optimal_w_alpha}, we turn the feature maps into the kernel in the primal space. Therefore, when it comes to updating $\bbeta$ and $ Q^{t}$, we do the minimization in the primal space.

\begin{Proposition}\label{pro:converge2}
Algorithm~\ref{algo:step2} converges to a critical point.
\end{Proposition}
\begin{IEEEproof}
See Appendix~\ref{prf:converge2}.
\end{IEEEproof}

\begin{algorithm}
\caption{Optimizing sensor decision rules and fusion center rules}
\label{algo:step2}
    \begin{algorithmic}[1]
    \STATE{\textbf{input:} $\{h_{i},g_{i},x^{1}_{i},\ldots,x^{s}_{i}\}_{i=1}^n$}
    \STATE{\textbf{initialization:} $w[0]\in\Real^n,\bbeta[0]\leftarrow\tilde{\bbeta},
     Q[0]\leftarrow\tilde{ Q},k=0$},
    \REPEAT
    \STATE{
        \begin{itemize}[leftmargin=*]
            \item $k=k+1$,
            \item Fix $\bbeta[k-1]$ and $ Q[k-1]$, and solve the following convex optimization problem,
                \begin{align*}
                    \balpha[k]=&\argmax_{\balpha\in\Real^n}\left\{F^*(\balpha,Q[k-1])\right\},
                \end{align*}
         and we obtain $w_H[k]=\sum^{n}_{i=1}\alpha_{i}[k]h_{i}\Phi_Q(\bx)$,
            \item Fix $w_H[k]$ and $ Q[k-1]$, update
                \begin{align*}
                     \bbeta[k]=\argmax_{\bbeta\in\Real^n}\ofrac{\mu}\log\left(\hat{R}^*(\bbeta, Q[k-1])-\theta\right),
                \end{align*}
            \item Fix $w_H[k]$, $\bbeta[k]$, and $ Q^1[k],\ldots, Q^{t-1}[k], Q^{t+1}[k-1], Q^s[k-1]$, update
                \begin{align*}
                     Q^t[k]=&\argmin_{ Q^t\in\calQ}F_0(w_H[k], Q^1[k],\ldots, Q^{t-1}[k],\\
                     & Q^t, Q^{t+1}[k-1], Q^s[k-1],\bbeta[k]),
                \end{align*}
        \end{itemize}
    }
    \UNTIL{$\frac{F_0(w_H[k-1],Q[k-1],\bbeta[k-1])-F_0(w_H[k],Q[k],\bbeta[k])}{F_0(w_H[k-1],Q[k-1],\bbeta[k-1])}\leq\epsilon$,}
    \STATE{\textbf{output:}} $\balpha[k],\bbeta[k], Q[k]$.
    \end{algorithmic}
\end{algorithm}

\subsection{Extension to \texorpdfstring{$m$}{m}-ary Private Hypothesis}\label{subsec:m-ary}
In this section, we extend our NPO framework to the case where $G$ is an $m$-ary hypothesis, with $m >2$. For simplicity, the public hypothesis $H$ remains as a binary hypothesis since in many IoT applications like intrusion detection, detection of whether an event has occurred is the main phenomenon of interest. By using standard multi-class classification techniques, our framework can be generalized to the case where $H$ is a multi hypothesis. We refer the reader to \cite{Lorena2009} for details.

Suppose that $G \in \calG=\{0,1,\ldots,m-1\}$, where $m>2$. For a detector $\gamma_g$ that distinguishes between the hypothesis pair $(0,g)$, where $1\leq g < m$, let
\begin{align*}
R_g(\gamma_g) &= \ofrac{2}\left(\P(\gamma_g(Z)=g\mid G=0)+\P(\gamma_g(Z)=0\mid G=g)\right),
\end{align*}
and
\begin{align}
\ell_g(\bz) &=\frac{p_{Z|G}(\bz|g)}{p_{Z|G}(\bz|0)},\label{ellg}
\end{align}
for $\bz\in\calZ^s$. The following result follows from Proposition~\ref{thm:PrivacyMetrics}\ref{it:infoprivacy-bayes}.

\begin{Theorem}\label{thm:m-empricalInfoPrivacy}
Suppose that $\min_{g\in\calG} p_G(g)> 0$ and the support $\calD$ of the conditional distributions of $Z$ given $G=g$ for all $g\in\calG$ are the same. If $\min_{1\leq g < m,\gamma_g} R_g(\gamma_g) \geq \theta$, with $\theta\in[0,1/2]$, then $G$ given $Z$ achieves $\epsilon$-information privacy where $\epsilon = 2\log \frac{c'}{(c'+2\theta-1)_+}$, with $c'=\min_{1\leq g < m}\{\allowbreak\P(\ell_g(\bz)=\min_{\bz\in\calD}\ell_g(\bz)\mid G=0),\allowbreak\P(\ell_g(\bz)=\max_{\bz\in\calD}\ell_g(\bz)\mid G=g)\} > 0$.
\end{Theorem}
\begin{IEEEproof}
See Appendix~\ref{prf:m-empricalInfoPrivacy}.
\end{IEEEproof}

To achieve $(\epsilon,\delta)$-information privacy for $G$ for $\delta\in (0,1]$, we impose the following $m-1$ empirical normalized risk constraints:
\begin{align}
\min_{w_g\in\scH}\hat{R}_{g,n}(w_g,Q) \geq \theta,\ \text{for $g=1,\ldots,m-1$},\label{maryconstraints}
\end{align}
where
\begin{align}
\hat{R}_{g,n}(w,Q)&=\ofrac{2}\sum_{g'\in\{0,g\}}\sum_{i\in\calS_{g',n}}\frac{\phi(g'_i\ip{w}{\Phi_Q(\bx_i)})}{|\calS_{g',n}|}+\frac{\lambda_n}{2}\norm{w}^2,
\end{align}
with $g'_i = -1$ if $g_i = 0$ and $g'_i = 1$ otherwise.
%<*tag:m-ary>
Note that here, we only need to consider the empirical normalized risk of confusing $G=g \geq 1$ with $G=0$, instead of every pair of hypothesis values. The intuition is the same as in $m$-ary hypothesis testing, where \eqref{ellg} are sufficient statistics.
%</tag:m-ary>
Indeed, it can be shown, using the same arguments as in Theorem~\ref{thm:empricalInfoPrivacy} that the following holds (we omit the proof here):

\begin{Theorem}\label{thm:empricalInfoPrivacy_mary}
Suppose the corresponding assumptions as in Theorem~\ref{thm:empricalInfoPrivacy} for $m$-ary hypothesis $G$ hold. Then for any $\delta\in(0,1]$, there exists $n_0$, such that for all $n\geq n_0$, $G$ achieves $(\epsilon,\delta)$-information privacy for any $Q\in\calQ$ satisfying \eqref{maryconstraints}, where $\epsilon=\ofrac{2}\log\frac{c'}{\left(c'-2a(\phi(0)-\theta+\delta)^{1/r}\right)_+}$, with $c'$ as defined in Theorem~\ref{thm:m-empricalInfoPrivacy}.
\end{Theorem}

We apply Algorithm~\ref{algo:step1} to each of the $m-1$ constraints in \eqref{maryconstraints} to obtain a threshold $\theta^*_g$ for each $g=1,\ldots,m-1$. Then, we set $\theta=p\cdot\min_{1\leq g < m} \theta^*_g$, for some $p\in(0,1)$. Finally, in Algorithm~\ref{algo:step2}, the objective function is replaced by
\begin{align*}
\min_{w_H\in\scH,\bbeta_g, Q\in\calQ}F(w_H, Q) -\frac{1}{\mu}\sum_{g=1}^{m-1} \log(\hat{R}^*_g(\bbeta_g, Q)-\theta),
\end{align*}
where $\hat{R}^*_g(\bbeta_g, Q)$ is the dual form of $\hat{R}_{g,n}(w_g,Q)$ (see \eqref{dualR}). It can be shown that Proposition~\ref{pro:converge2} still holds.

\section{Simulations and Experiments}
\label{sec:simulation}

In this section, we first perform simulations to provide insights into how different parameters impact the performance of our NPO approach. We then test our algorithm on real datasets from the UCI Repository \cite{Lichman:2013}, and compare its performance with RUCA\cite{al2017ratio} and MDR\cite{diamantaras2016data}.

For simplicity, we use the \emph{count kernel} in our simulations, which is defined as $\kappa(\bz,\tilde{\bz})=\sum^{s}_{t=1}\indicator{z^{t}=\tilde{z}^{t}}$. Then, for any $Q\in\calQ$, we have
\begin{align*}
\kappa_{Q}(\bx,\tilde{\bx})
%=&\ip{\Phi_Q(\bx)}{\Phi_Q(\tilde{\bx})}\\
%=&\sum_{\bz\in\calZ^s}\sum_{\tilde{\bz}\in\calZ^s}Q(\bz\mid\bx)Q(\tilde{\bz}\mid\tilde{\bx})\sum^{s}_{t=1}\indicator{z^{t}=\tilde{z}^{t}}\\
=&\sum^{s}_{t=1}\sum_{z\in\calZ} Q^t(z\mid x^{t})Q^t(z\mid\tilde{x}^{t}),
\end{align*}
which can be computed with a time complexity of $\calO(s|\calZ|)$. On the other hand, for some kernels like the Gaussian kernel, the same computation incurs a time complexity of $\calO(|\calZ|^{2s})$. Note however that the count kernel does not satisfy Assumption~\ref{assumpt:phi2bayes}\ref{it:strictly positive}. Nevertheless, our simulations suggest that using the count kernel does not prevent our NPO approach from protecting the information privacy of $G$. Since many IoT devices and gateways that serve as fusion centers are embedded platforms with limited computation power \cite{sun2005distributed}, the count kernel allows practical implementation. Therefore, in our simulations, we evaluate the performance of our algorithms using the count kernel.

We choose the logistic loss function as $\phi$ in our simulations, and employ gradient descent in the optimization steps in Algorithms \ref{algo:step1} and \ref{algo:step2}.
%For example, the gradient when updating $Q^t$ in Algorithm~\ref{algo:step2} is
%\begin{align}
%&\frac{\partial F_{0}(w_H,\bbeta, Q)}{\partial Q^{t}(\bar{z}^t|\bar{x}^t)}\nonumber\\
%%=&\sum_{i=1}^n\frac{\partial\phi(b)}{\partial b}\ip{w_H}{\sum_{z=1}^s\frac{Q(\bz|\bx_{i})}{
%%Q^{t}(z^{t}|x_{i}^t)}\Phi(\bz)}\indicator{z^t=\bar{z}^t}\indicator{x_{i}^t=\bar{x}^t}\nonumber\\
%%&+\ofrac{2\mu \blue{d\lambda_n}}\sum_{i=1}^n\sum_{j=1}^n\beta_{i}\beta_{j}g_{i}g_{j}\frac{\partial\kappa_{Q}(\bx_{i},\bx_{j})}{\partial Q^{t}(\bar{z}^t|\bar{x}^t)}\nonumber\\
%=&\frac{\partial\phi(b)}{\partial b}\sum_{i,j,\bz,\bz'}\alpha_{j}h_{j}\frac{Q(\bz'|\bx_j)Q(\bz|\bx_i)}{Q^{t}(z^{t}|x_{i}^t)}
%\kappa(\bz,\bz')\indicator{z^t=\bar{z}^t}\indicator{x_{i}^t=\bar{x}^t}\nonumber\\
%&+\ofrac{2\mu \blue{d\lambda_n}}\sum_{i,j,\bz,\bz'}\beta_{i}\beta_{j}g_{i}g_{j}\kappa(\bz,\bz')\frac{Q(\bz|\bx_i)Q(\bz'|\bx_j)}{Q^t(z^t|x^t)}(\indicator{z^t=\bar{z}^t}\nonumber\\
%&+\indicator{z^{\prime t}=\bar{z}^t})\indicator{x_{i}^t=\bar{x}^t},\label{eq:gradientQ}
%\end{align}
%with $b=h_{i}\ip{w_H}{\Phi_Q(\bx_i)}$, $\blue{d}=\hat{R}^*(\bbeta,Q)-\theta$.
The complexity of both Algorithms \ref{algo:step1} and \ref{algo:step2} using gradient descent in each optimization step and the count kernel is $\calO(n^2s|\calZ|)$. We use $\Delta_1=\Delta_2=0.005$ in Algorithm~\ref{algo:step1} throughout.

\subsection{Synthetic Data Set}
\label{subsec:synthetic}
In this section, we generate a synthetic data set to verify the performance of our proposed method. We first consider the case where $G$ has uniform prior probability and then give an example where $G$ has a skewed prior. Finally, we present simulation results for the case where $G$ is a $m$-ary hypothesis, with $m>2$. To evaluate the performance of our algorithm, we compute the Bayes errors for detecting $H$ and $G$ since these are the minimum detection errors any detector can achieve so that our results are oblivious to the choice of learning method adopted by the fusion center. We also show the Bayes errors of detecting $H$ and $G$ when $Z=X$ (i.e., the raw sensor observations are available at the fusion center) as a baseline for comparison.

\subsubsection{Performance of NPO framework}
Consider a network of $4$ sensors and a fusion center. Each sensor observation $x^t_i \in \calX=\{1,2,\ldots,8\}$ is generated according to Table~\ref{tab:discreteX}, where $n_i^t$ is distributed uniformly over $\{-1,0,+1\}$. In this set of simulations, $G$ has uniform prior. Conditioned on $(H,G)$, sensor observations are independent of each other. We generate $80$ i.i.d.\ training samples and $1000$ i.i.d.\ testing samples.

\begin{table}[!ht]
    \caption{Sensor observations $X$ for different realizations of $(H,G)$}
	\centering
		\begin{tabular}{ | c | c |  }\hline
    		$(h_i,g_i)$ 			& $x^t_i$\\ \hline 									
                 $(-1,-1)$			& $2+n_i^t$ 	\\ \hline
    		$(-1,1)$			& $4+n_i^t$ 	\\ \hline
            $(1,-1)$			& $6+n_i^t$ 	\\ \hline
    		$(1,1)$				& $8+n_i^t$ 	\\ \hline
		\end{tabular}
	\label{tab:discreteX}
\end{table}

\begin{figure}[!ht]
    \centering
    \includegraphics[width=0.5\textwidth]{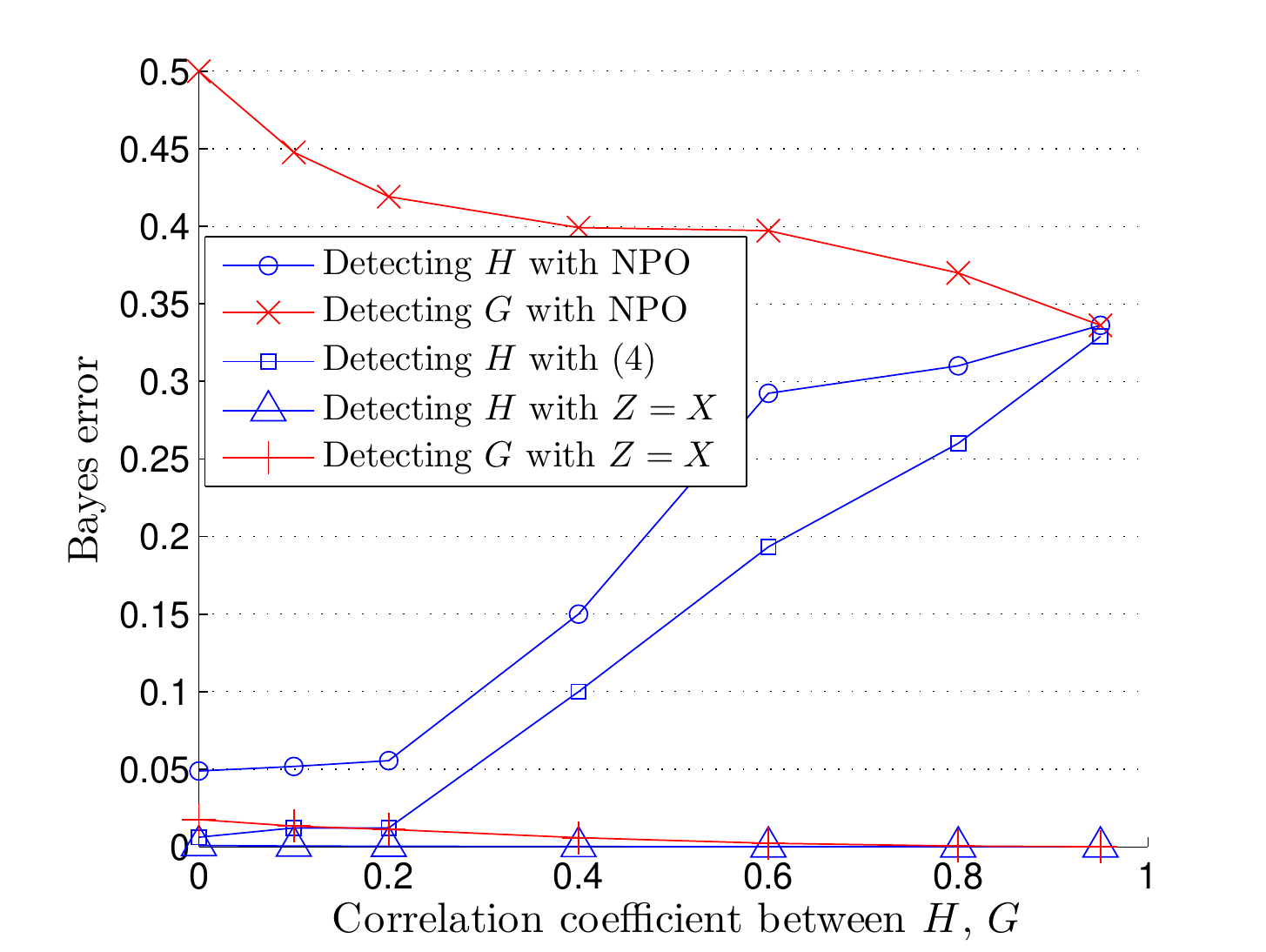}
    \caption{Error rate of detecting $H$ and $G$ with different correlation coefficient between $H$ and $G$.}
    \label{fig:correlation}
\end{figure}

In Fig.~\ref{fig:correlation}, we show how the correlation coefficient between $H$ and $G$ affects the detection error rates. Recall that the privacy threshold is chosen to be $\theta = p\theta^*$, where $\theta^*$ is found using Algorithm~\ref{algo:step1}. We use a privacy threshold ratio $p=0.999$. We generated both $H$ and $G$ with zero mean, but varying correlation coefficient between them. As the correlation between $H$ and $G$ becomes larger, the Bayes error for both hypotheses converge to each other. As expected, to achieve a reasonable Bayes error for $H$ requires that $H$ is not too correlated with $G$.

%<*correlation>
We compare the performance of our approach with the optimal detectors found using \eqref{gop}, which assumes knowledge of the underlying joint distribution. We use the same $\theta$ as that in our NPO to achieve the same error rate for $G$. We see that our NPO achieves an error rate for $H$ that is not too different from the optimal detectors, even though no prior knowledge of the underlying distribution is assumed in NPO.
%</correlation>

\begin{figure}[!ht]
    \centering
    \includegraphics[width=0.5\textwidth]{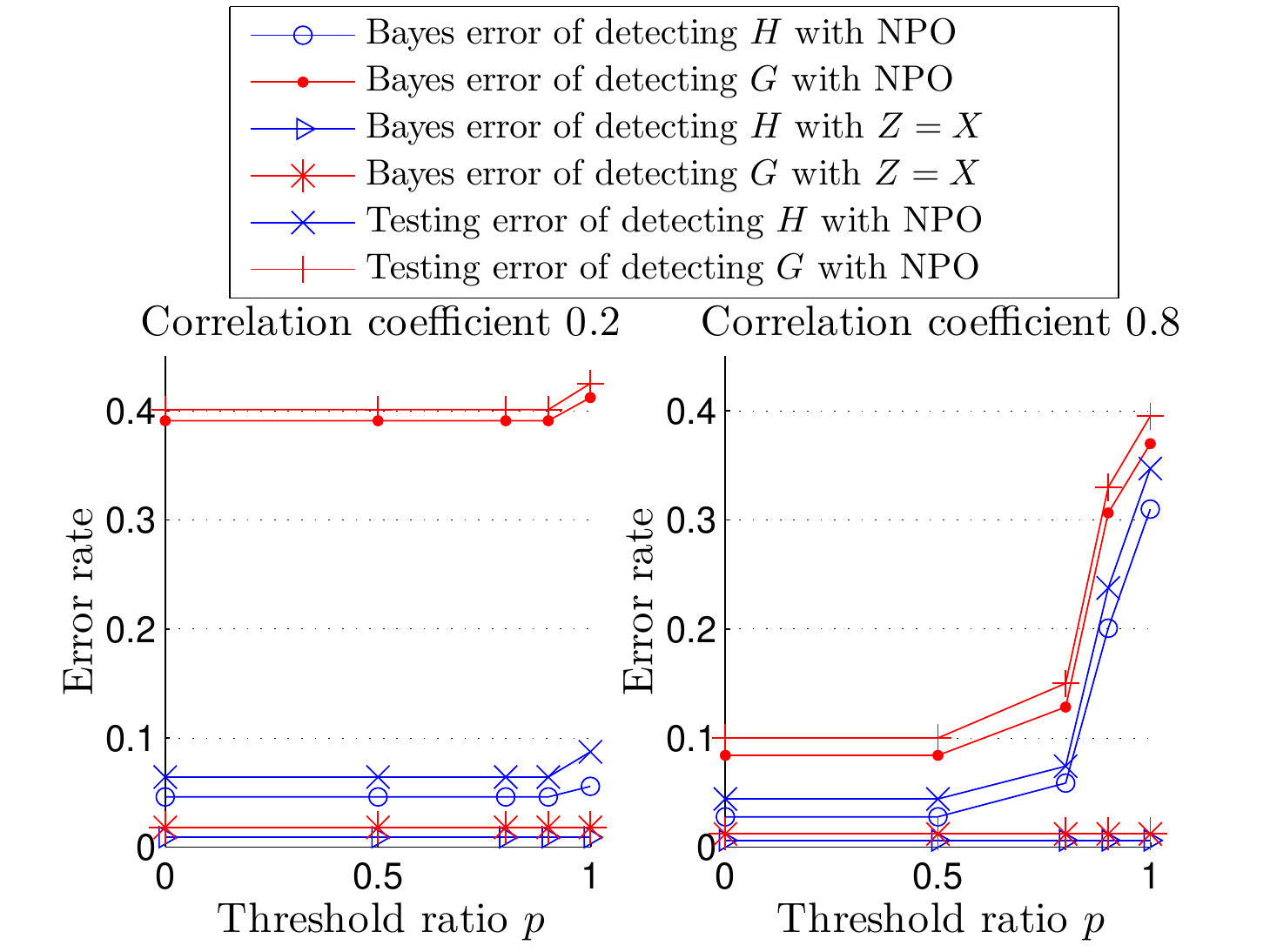}
    \caption{Error rate of detecting $H$ and $G$ with varying privacy threshold ratio $p$.}
    \label{fig:thresholdH}
\end{figure}

In Fig.~\ref{fig:thresholdH}, we show the effect of the threshold ratio $p$ on the error rates of detecting $H$ and $G$, when their correlation coefficients are $0.2$ and $0.8$, respectively. The testing error is the empirical detection rate of the classifier found using Algorithm~\ref{algo:step2} applied to the testing samples we generated.
%<*threshold>
We see that when the correlation between $H$ and $G$ is small, $p$ has no significant effect on their error rates over a large range. This is because the privacy mapping $Q$ that minimizes the error rate of $H$ does not contain much information about $G$. We also observe that with the NPO framework, the error rate for $H$ is not significantly higher than using the raw sensor observations, whereas the error rate for $G$ is increased significantly. However when the correlation between $H$ and $G$ is large, $p$ has significant impact on the error rates. In this case, as expected, we cannot find a $p$ that induces a low error rate for $H$ and a high error rate for $G$.
%</threshold>

%In the previous simulations, we assume an i.i.d.\ model in which sensor observations are conditionally independent of each other. We next consider a chain model in which each sensor's observation is defined as $X_{\text{chain}}^{t}=(X^{t-1}+X^{t}+X^{t+1})/3$, where the $X^t$ are generated according to the i.i.d.\ model, with $X^t=0$ if $t<1$ or $t>4$. The correlation coefficient of $H$ and $G$ is chosen to be $0.08$, and we vary the size of the decision space $\calZ$ to evaluate the performance of our algorithms. As shown in Fig.~\ref{fig:Quantization}, the testing error for detecting $H$ is much smaller than that for $G$.
%%<*reason>
%However, we see that the chain model yields a worse result because it is more difficult for local privacy mappings to remove information about $G$ while retaining that for $H$, as the correlation between sensor observations increases.
%%</reason>

%\begin{figure}[!ht]
%    \centering
%    \includegraphics[width=0.5\textwidth]{chain}
%    \caption{Testing error of detecting $H$ and $G$ with varying $|\calZ|$.}
%    \label{fig:Quantization}
%\end{figure}

\subsubsection{Comparison with the empirical Bayes error privacy metric}
\label{subsubsec:compare}
Example~\ref{eg:Bayes vs InfoPrivacy} shows that even if the Bayes error is large, information privacy cannot be guaranteed. In this simulation, we demonstrate that using the empirical risk corresponding to the Bayes error \cite{SunTay:C16} as a privacy metric can lead to privacy mappings that result in a $p_{G\mid Z}(g\mid \bz)$ that is significantly different from $p_G(g)$, i.e., some realization $Z=\bz$ can leak significant information about the true state of $G$. With the empirical risk corresponding to the Bayes error, our optimization problem becomes:
\begin{align}
\begin{aligned}\label{old method}
    &\min_{w_H\in\scH,Q\in\calQ} F(w_H,Q)\\
\text{s.t.\ }&\min_{w_G\in\scH}\ofrac{n}\sum_{i=1}^n\phi\left(g_{i}\ip{w_G}{\Phi_Q(\bx_i)}\right)+\frac{\lambda}{2}\norm{w_G}^{2}\geq\theta.
\end{aligned}
\end{align}
We let $\P(G=1)=0.95$ and the correlation coefficient between $H$ and $G$ to be $0.23$. We use a similar procedure as Algorithms \ref{algo:step1} and \ref{algo:step2} to find a privacy threshold and the sensor privacy mappings for \eqref{old method}. The results are shown in Table~\ref{tab:old method}. We see that the Bayes errors of detecting $H$ and $G$ respectively are similar for both the NPO approach and \eqref{old method}. However, NPO achieves a much smaller privacy budget $\epsilon$ (i.e., stronger information privacy) than \eqref{old method}.

\begin{table*}[!hbt]
\centering
\caption{Comparison of NPO with \eqref{old method}}
\label{tab:old method}
\begin{tabular}{|c|c|c|c|c|}
\hline
   & Bayes Error $H$ & Bayes Error $G$ & $R(\gamma_G)$ & Information privacy budget, $\epsilon$ \\ \hline
 NPO  	& $0.03$ & $0.05$ & $0.38$ & $1.08$ \\ \hline
(\ref{old method}) 			& $0.02$ & $0.05$ & $0.24$ & $ 3.91$ \\ \hline
\end{tabular}
\end{table*}

\subsubsection{NPO with \texorpdfstring{$m$}{m}-ary private hypothesis}\label{subsubsec:m}
We next show the performance of NPO when $G$ is a $m$-ary hypothesis with $m >2$. We let $(H,G) \in \{-1,1\}\times\{0,\ldots,m-1\}$, $H$ and $G$ be uncorrelated, with both having uniform priors. We let $X^t= m(H+1)+2(G+1)+N^t$, where $t=1,2,3,4$, and for sensors $t=5,\ldots,4+m$, we let
\begin{align*}
%=$ if $$, and $X^t=0$ otherwise.
X^t=\left\{
                \begin{array}{ll}
                  H+N^t,&\text{if } G-H/2-0.5\pmod m\equiv t-5\\
                  0,& \text{otherwise}.
                \end{array}
              \right.
\end{align*}
The noise $N^t$ is chosen uniformly at random from $\{-1,0,+1\}$. We perform training using $80$ training samples, with $\calZ=\{1,2\}$.

We compare NPO with two other methods: (i) detecting $H$ and $G$ when $Z=X$; and (ii) finding the sensor local decision rules by minimizing $F(w_H,Q)$ without any privacy constraints. We call this the nonparametric decentralized detection (NDD) method \cite{NguWaiJor:05}.

We see from Table~\ref{tab:mary} that with NPO, the Bayes error of detecting $H$ is small, while the Bayes error of detecting $G$ is large, and increasing with $m$. Note that the maximum possible Bayes error for $G$ (achieved by random guessing, i.e., $Z$ is independent of $G$) is $1-1/m$. We see that our comparison methods fail to protect the privacy of $G$.

\begin{table*}[hbt!]
\centering
\caption{Bayes Error with $m$-ary Private Hypothesis}
\label{tab:mary}
\begin{tabular}{|c|c|c|c|c|c|c|}
\hline
$m$ & \begin{tabular}[c]{@{}c@{}}Detect $H$\\ with NPO\end{tabular}        & \begin{tabular}[c]{@{}c@{}}Detect $G$\\ with NPO\end{tabular} &\begin{tabular}[c]{@{}c@{}}Detect $H$\\with $Z=X$\end{tabular}&\begin{tabular}[c]{@{}c@{}}Detect $G$\\ with $Z=X$\end{tabular} & \begin{tabular}[c]{@{}c@{}}Detect $H$\\ with NDD\end{tabular} &\begin{tabular}[c]{@{}c@{}}Detect $G$\\ with NDD \end{tabular}   \\ \hline
$3$ & $3.62\times 10^{-2}$ & $0.65$  &  $3.77\times 10^{-3}$ & $1.21\times 10^{-2}$ & $2.41\times 10^{-2}$ & $0.13$\\\hline
$4$ & $2.72\times 10^{-2}$ & $0.73$  & $1.10\times 10^{-3}$ & $ 3.92\times 10^{-3}$ & $1.85\times 10^{-2}$ & $0.19$\\\hline
$5$ & $2.13\times 10^{-2}$   & $0.79$ &$ 3.02\times 10^{-3}$ &$4.22\times 10^{-3}$ & $1.78\times 10^{-2}$ & $0.23$\\ \hline
\end{tabular}
\end{table*}

\subsection{UCI Repository Data Sets}\label{subsec:dataset}

We test our algorithm on the OPPORTUNITY Activity Recognition Data Set \cite{Chavarriaga2013} and Census
(Adult) Data Set (both available at UCI Repository\cite{Lichman:2013}), and compare its performance with RUCA\cite{al2017ratio} and MDR\cite{diamantaras2016data}. In our NPO, we set the local decision space of each sensor to be $\calZ=\{1,2\}$.

\subsubsection{Data and implementation}
The OPPORTUNITY Activity Recognition Data Set consists of recorded readings of on-body, object and ambient sensors installed in a living space, while a person executes typical daily activities. We let the public hypothesis be locomotion detection (standing or walking), and the private hypothesis be the object the person is touching (a drawer or a dish washer). We used the data from `S2-Drill', and Weka \cite{Hall2009} to select $s=15$ sensors that are most correlated with our chosen labels. Since the sensor reading is continuous, unsupervised discretization was applied to quantize each continuous sensor reading to $10$ levels. We randomly sampled $n=40$ instances of training data, and $3427$ instances of testing data.

The Census (Adult) Data Set contains information about a population, such as income, marital status, gender, and nationality. We let the public hypothesis be whether a person has high income ($>50$K) or low income ($\leq 50$K), and the private hypothesis be the marital status of the person. Following \cite{al2017ratio}, we group `Married-civ-spouse', `Married-spouse-absent' and `Married-AFspouse' into a single category called `Married', while `Divorced', `Separated' and `Widowed' are grouped into a single category called `Used to be Married'. The `Never Married' category was left as it is. Thus the private hypothesis is a $3$-ary hypothesis. We used data from `adult.data', and selected $s=5$ attributes (age, education, occupation, relationship, sex) that are most correlated with our chosen labels. Unsupervised discretization is applied to quantize each numerical attributes to $15$ levels. We then randomly sampled $n=180$ instances of training data, and $2100$ instances of testing data.

\subsubsection{Results and discussion}
%<*first>
To the best of our knowledge, our work is the first to provide a nonparametric approach to design \emph{decentralized} sensor privacy mappings. Therefore, we compare our method to RUCA\cite{al2017ratio} and MDR\cite{diamantaras2016data}, which are nonparametric approaches that require a trusted data curator to first aggregate data from all sensors before applying an overall privacy mapping. Parameter settings for RUCA and MDR are as proposed in \cite{al2017ratio} and \cite{diamantaras2016data}, respectively. We also compare with the case where sensors send their observations directly to the fusion center, namely $Z=X$.
%</first>

To estimate the information privacy budget achieved by each method, we compute
\begin{align*}
\hat{\epsilon} = \max\limits_{g\in\calG,\bz\in\calZ^s} \left|\log \frac{\hat{p}_{G,Z}(g,\bz)}{\hat{p}_G(g)\hat{p}_Z(\bz)}\right|,
\end{align*}
where $\hat{p}_{A}(a)$ is the empirical probability of the event $\{A=a\}$. Note that a smaller $\hat{\epsilon}$ implies stronger information privacy.

From Table~\ref{tab:activity data set} and Table~\ref{tab:census}, we observe that our proposed NPO is comparable to the other nonparametric methods that need a trusted data curator. We also observe that NPO achieves stronger information privacy than the other methods, since it explicitly optimizes for this criterion.

\begin{table}[!ht]
\centering
\caption{Detection errors using the OPPORTUNITY Activity Recognition Data Set.}
\label{tab:activity data set}
\begin{tabular}{|c|c|c|c|}
\hline
Detection Method  &  \begin{tabular}[c]{@{}c@{}}$H$\\ Locomotion\end{tabular} & \begin{tabular}[c]{@{}c@{}}$G$\\Object\end{tabular}&$\hat{\epsilon}$  \\ \hline
NPO                  & $10.53\%$       & $43.17\%$  &$0.47$\\ \hline
MDR                & $12.56\%$       & $40.16\%$ &    \multirow{6}{*}{$0.69$}   \\ \cline{1-3}
DCA                & $10.88\%$       & $42.62\%$ &        \\ \cline{1-3}
RUCA ($\rho_p=1$)     & $10.23\%$       & $45.73\%$ &       \\ \cline{1-3}
RUCA ($\rho_p=100$)     & $10.10\%$       & $43.01\%$ &       \\ \cline{1-3}
RUCA ($\rho_p=1000$)     & $10.10\%$       & $43.78\%$ &       \\ \cline{1-3}
$Z=X$ &$10.05\%$ & $5.57\%$ &\\ \hline
\end{tabular}
\end{table}

\begin{table}[!ht]
\centering
\caption{Detection errors using the Census (Adult) Data Set.}
\label{tab:census}
\begin{tabular}{|c|c|c|c|}
\hline
Detection Method  &  \begin{tabular}[c]{@{}c@{}}$H$\\ Income\end{tabular} & \begin{tabular}[c]{@{}c@{}}$G$\\Marital-status\end{tabular}& $\hat{\epsilon}$   \\ \hline
NPO                  & $37.67\%$       & $62.14\%$   &   $0.82$  \\ \hline
MDR                & $37.57\%$       & $64.02\%$     & \multirow{6}{*}{$1.61$}  \\ \cline{1-3}
DCA                & $38.38\%$       & $56.33\%$     &    \\ \cline{1-3}
RUCA ($\rho_p=1$)     & $41.24\%$       & $65.25\%$     &   \\ \cline{1-3}
RUCA ($\rho_p=100$)     & $41.10\%$       & $64.14\%$   &     \\ \cline{1-3}
RUCA ($\rho_p=1000$)     & $40.67\%$       & $65.86\%$  &      \\ \cline{1-3}
$Z=X$ &$34.05\%$ & $30.48\%$  & \\ \hline
\end{tabular}
\end{table}

\section{Conclusions}\label{sec:conclusions}

Information privacy is the protection against statistical inference of a private phenomenon. In this paper, we proved that for a binary hypothesis, ensuring that the average of its Type I and II errors is large, is sufficient to guarantee information privacy. We proposed a nonparametric privacy-aware optimization framework to find sensor privacy mappings that can limit the fusion center's detection rate of a private hypothesis, without significantly compromising the detecting ability of the public hypothesis. We proved a theoretical guarantee of weak information privacy for our proposed framework, and proposed iterative algorithms to solve the proposed optimization problem. Our approach is suitable for IoT networks in which every sensor operates independently of the others.

Future research directions include generalizing our privacy-aware optimization framework to networks with multiple layers so that nodes of different sensing or computation ability can apply local decision rules of different complexities to achieve a better overall utility-privacy tradeoff. It would also be of interest to design sensor privacy mappings that achieve both information and data privacy simultaneously.

\begin{appendices}
\section{Proof of Proposition~\ref{thm:PrivacyMetrics}}\label{prf:PrivacyMetrics}
We first show~\ref{it:bayes-infoprivacy}. Let $P_{e} = \P(\gamma_G(Z)\neq G)$ and $\calI(\cdot;\cdot)$ denote the mutual information operator. From Fano's inequality \cite{Cover2006}, we have
\begin{align*}%\label{eq:fano's inequality}
    \calH(P_{e}) &\geq \calH(G \mid Z)\\
    &= \calH(G)-\calI(G;Z)\\
    &= \calH(G)-\E[\log\frac{p_{G|Z}(G|Z)}{p_{G}(G)}]\\
    &\geq \calH(G) - \epsilon,
\end{align*}
where the last inequality follows from Definition~\ref{def:info privacy}. Since $\calH(P_{e})$ is a non-decreasing function of $P_{e}$ for $0\leq P_{e} \leq 1/2$, we obtain $P_e\geq \theta$, where $\theta$ satisfies \eqref{ineq:delta-infoprivacy}. Part~\ref{it:bayes-infoprivacy} now follows from $R(\gamma_G) \geq P_e / (2\max_g p_G(g))$.

We next show~\ref{it:infoprivacy-bayes}. Let $\gamma_G(Z) = \argmin_{\gamma} R(\gamma)$. It can be shown that $\gamma_G(Z) = 1$ if $\ell(Z) \geq 1$ and $\gamma_G(Z)=-1$ otherwise. Let $\Gamma = \{ \bz \in \calD : \ell(\bz) \geq 1\}$ and $\Gamma^c = \calD \backslash \Gamma$. We have
\begin{align}
    R(\gamma_G) &= \ofrac{2} - \ofrac{2} \sum_{\bz\in\Gamma} \left(p_{Z\mid G}(\bz \mid 1) - p_{Z\mid G}(\bz \mid -1)\right)\label{eq:R1}\\
    &=  \ofrac{2} + \ofrac{2} \sum_{\bz\in\Gamma^c} \left(p_{Z\mid G}(\bz \mid 1) - p_{Z\mid G}(\bz \mid -1)\right). \label{eq:R2}
\end{align}
For any $\bz' \in \Gamma$, if $R(\gamma_G) \geq \theta$, we have from \eqref{eq:R1},
\begin{align}
    1 - 2\theta
    &\geq \sum_{\bz\in\Gamma} \left(p_{Z\mid G}(\bz \mid 1) - p_{Z\mid G}(\bz \mid -1)\right)\nonumber\\
    &=\sum_{\bz\in\Gamma} p_{Z\mid G}(\bz \mid 1)\left(1 - \ofrac{\ell(\bz)}\right)\nonumber\\
    &\geq  \left(1 - \ofrac{\ell(\bz')}\right)\sum_{\{\bz: \ell(\bz) \geq \ell(\bz')\}} p_{Z\mid G}(\bz \mid 1) \nonumber\\
		& \geq \left(1 - \ofrac{\ell(\bz')}\right) \P(\ell(\bz)=\max_{\bz\in\calD}\ell(\bz)){G=1} \nonumber\\
    & \geq c \left(1 - \ofrac{\ell(\bz')}\right). \label{ineq:zinGamma}
\end{align}
From \eqref{ineq:zinGamma}, we obtain
\begin{align*}
\ofrac{\ell(\bz')} \geq 1 - \frac{1-2\theta}{c},
\end{align*}
and
\begin{align}
    1 \leq \ell(\bz') \leq \ofrac{(1- (1-2\theta)/c)_+} = e^{\epsilon}. \label{zinGammaBdd}
\end{align}
Similarly, for any $\bz' \in \Gamma^c$, a similar argument using \eqref{eq:R2} yields
\begin{align}
    e^{-\epsilon} \leq \ell(\bz') \leq 1. \label{znotinGammaBdd}
\end{align}
Combining \eqref{zinGammaBdd} and \eqref{znotinGammaBdd}, we have for any $\bz \in \calD$,
\begin{align}\label{ineq:ellzbounds}
    e^{-\epsilon} \leq \ell(\bz) \leq  e^{\epsilon},
\end{align}
and for any $g\in \{-1,1\}$,
\begin{align}\label{pZbounds}
\begin{aligned}
    e^{-\epsilon}p_{Z\mid G}(\bz \mid g)& \leq p_{Z}(\bz) \\
    &= \E{G}[p_{Z\mid G}(\bz\mid G)] \leq  e^{\epsilon}p_{Z\mid G}(\bz \mid g).
    \end{aligned}
\end{align}
We then obtain \eqref{ineq:info privacy} by noting that
\begin{align*}
    \frac{p_{G \mid Z}( g \mid \bz)}{p_{G}(g)} = \frac{p_{Z \mid G}( \bz \mid g)}{p_{Z}(\bz)},
\end{align*}
and the proof of the theorem is complete.

\section{Proof of Theorem~\ref{thm:empricalInfoPrivacy}}\label{prf:empricalInfoPrivacy}

We start with the following elementary lemma.
\begin{Lemma_A}\label{lemma:leaveoneout}
For any $w\in\scH$,
\begin{align*}
\E[\hat{R}_n(w,Q)]=R_\phi(w,Q)+\frac{\lambda_n}{2}\norm{w}^2,
\end{align*}
where the expectation is taken with respect to the training samples $(X_i,G_i)_{i=1}^n$.
\end{Lemma_A}
\begin{IEEEproof}
For each $g\in\{-1,1\}$, we have
\begin{align*}
    &\E[\sum_{i\in\calS_{g,n}}\frac{\phi\left(G_i\ip{w}{\Phi_Q(X_i)}\right)}{|\calS_{g,n}|}]\\
    =&\E[\E{X}[\sum_{i\in\calS_{g,n}}\frac{\phi\left(G_i\ip{w}{\Phi_Q(X_i)}\right)}{|\calS_{g,n}|}]{ G_1,\ldots,G_n}]\\
    =&\E[\ofrac{|\calS_{g,n}|}\sum_{i\in\calS_{g,n}}\E{X}[\phi\left(G_i\ip{w}{\Phi_Q(X_i)}\right)]{G_i=g}]\\
    =&\E{X}\left[\phi\left(g\ip{w}{\Phi_Q(X)}\right)\mid G=g\right].%\label{eq:expectation}
\end{align*}
The lemma now follows from \eqref{FG} and \eqref{Rphi}.
\end{IEEEproof}

We now prove Theorem~\ref{thm:empricalInfoPrivacy}. Let $\hat{w}_{G,n}=\displaystyle\argmin_{w\in\scH} \hat{R}_n(w,Q)$. From Markov's inequality, we have
%<*markov>
\begin{align}
    &\P\left(\hat{R}_n(\hat{w}_{G,n},Q)\geq\inf_{w\in\scH}R_\phi(w,Q)+\delta\right)\nonumber\\
    &\leq \ofrac{\delta}\left(\E[\hat{R}_n(\hat{w}_{G,n},Q)]-\inf_{w\in\scH}R_\phi(w,Q)\right)\nonumber\\
    &=\ofrac{\delta}\left(\E[\min_{w\in\scH}\hat{R}_n(w,Q)]-\inf_{w\in\scH}R_\phi(w,Q)\right)\nonumber\\
    &\leq \ofrac{\delta}\left(\inf_{w\in\scH}\E[\hat{R}_n(w,Q)]-\inf_{w\in\scH}R_\phi(w,Q)\right)\nonumber\\
    &=\ofrac{\delta}\left(\inf_{w\in\scH}\{R_\phi(w,Q)+\frac{\lambda_n}{2}\norm{w}^2\}-\inf_{w\in\scH}R_\phi(w,Q)\right),\label{eq:gotozero}
\end{align}
%</markov>
where the last equality follows from Lemma~\ref{lemma:leaveoneout}. If $n\to 0$ in \eqref{eq:gotozero}, its right hand side approaches to $0$ uniformly in $Q$ because of Assumption \ref{assumpt:phi2bayes}\ref{it:assumpt} and $\calQ$ is a finite set. Therefore,
%<*markov2>
there exists $n_0$, such that for all $n\geq n_0$, we have with probability $1-\delta$,
\begin{align}\label{eq:R}
    \hat{R}_n(\hat{w}_{G,n},Q)\leq&\inf_{w\in\scH}R_\phi(w,Q)+\delta.
\end{align}
%</markov2>

Let $\widetilde{G}$ be a binary hypothesis with uniform prior, and satisfies $p_{Z|\widetilde{G}}(\bz|g)=p_{Z|G}(\bz|g)$ for all $\bz\in\calZ^s$ and $g\in\{-1,1\}$. We then have
\begin{align}
    R_\phi(w,Q)=&\E[\phi\left(\widetilde{G}\ip{w}{\Phi_{Q}(X)}\right)]\nonumber\\
    =&\E[\phi\left(\widetilde{G}\ip{w}{\E{Z}[\Phi(Z)]{X}}\right)]\nonumber\\
    \leq & \E[\E{Z}[\phi\left(\widetilde{G}\ip{w}{\Phi(Z)}\right)]{X}]\nonumber\\
    =&\E[\phi\left(\widetilde{G}\cdot\gamma(Z)\right)]\nonumber\\
    =&\E[\left(\eta(Z)\phi(\gamma(Z))+(1-\eta(Z))\phi(-\gamma(Z))\right)],\label{2risk}
\end{align}
where the inequality follows from Jensen's inequality, and we let $\gamma(\bz)=\ip{w}{\Phi(\bz)}$ in the penultimate equality. Recall that $\eta(\bz) = \P(\widetilde{G}=1){Z=\bz}$. For each $\bz \in \calZ^s$, let $\gamma^*(\bz)=\arginf_{\gamma\in\Real}\left(\eta(\bz)\phi(\gamma)+(1-\eta(\bz))\phi(-\gamma)\right)$, and let $K$ be the Gram matrix of $\kappa(\cdot,\cdot)$ with respect to all elements in $\calZ^s = \{\bz_1,\ldots,\bz_{|\calZ^s|}\}$. From Assumption~\ref{assumpt:phi2bayes}\ref{it:strictly positive}, since $K$ is strictly positive definite, we can define
\begin{align*}
    [\psi_1,\psi_2,\ldots,\psi_{|\calZ^s|}]=[\gamma^*(\bz_1),\gamma^*(\bz_2),\ldots,\gamma^*(\bz_{|\calZ^s|})]\cdot\mathit{K}^{-1},
\end{align*}
and take $\gamma(\bz)=\sum_{i=1}^{|\calZ^s|} \psi_i \kappa(\bz_i,\bz) = \gamma^*(\bz)$ for all $\bz\in\calZ^s$. Therefore, from \eqref{2risk}, we obtain
\begin{align}
    \inf_{w\in\scH}R_\phi(w,Q)&\leq\E{Z}[R_\phi^*(\eta(Z))].\label{2minimum}
\end{align}
For $\eta\in[0,1]$, let $R^*(\eta)=1/2-|1/2-\eta|$. We then have
\begin{align}
    &\min_{\gamma}R(\gamma)\nonumber\\
    &=\min_{\gamma} \E[(1-\eta(Z))\indicator{\gamma(Z)\geq 0}+\eta(Z)\indicator{\gamma(Z)<0}]\nonumber\\
    &=\ofrac{2}-\E[\left|\ofrac{2}-\eta(Z)\right|]\nonumber\\
    &=\E[R^*(\eta(Z))].\label{Rstar}
\end{align}
From Assumption~\ref{assumpt:bound}, we obtain
\begin{align}
    \E[R_\phi^*(\eta(Z))]\leq&\phi(0)-\ofrac{a^r}\E[\left(\ofrac{2}-R^*\left(\eta(Z)\right)\right)^r]\nonumber\\
    \leq&\phi(0)-\ofrac{a^r}\left(\ofrac{2}-\E[R^*\left(\eta(Z)\right)]\right)^r,\label{ERphi}
\end{align}
where the last inequality follows from Jensen's inequality. Substituting \eqref{Rstar} into \eqref{ERphi}, we have
\begin{align*}
    \min_{\gamma}R(\gamma)\geq&\ofrac{2}-a\left(\phi(0)-\E[R_\phi^*\left(\eta(Z)\right)]\right)^\ofrac{r}\\
    \geq&\ofrac{2}-a\left(\phi(0)-\theta+\delta\right)^\ofrac{r},
\end{align*}
where the last inequality follows from \eqref{eq:R} and \eqref{2minimum}, and holds with probability $1-\delta$. The theorem then follows from Proposition~\ref{thm:PrivacyMetrics}, and the proof is complete.

\section{Proof of Proposition~\ref{pro:converge1}}\label{prf:converge1}

We first show that $-\hat{R}^*(\bbeta,  Q)$ is a strictly quasi-convex function of $Q$ if $\bbeta\ne 0$. The definition of strictly quasi-convexity is borrowed from \cite{Grippo2000} as follows.

\begin{Definition_A}\label{def:quasi-convex}
A function $f(x_{1},x_{2},\ldots,x_{s})$ is called a strictly quasi-convex function of $x_{1},\ldots,x_{s}$, if for all $k\in(1,\ldots,s)$,
\begin{align*}
f(x_1,&\ldots,p x_k^1+(1-p)x_k^2,\ldots,x_s)\\
&<\max(f(x_1,\ldots,x_k^1,\ldots,x_s),f(x_1,\ldots,x_k^2,\ldots,x_s)),
\end{align*}
for any $p\in (0,1)$.
\end{Definition_A}

\begin{Lemma_A}
Suppose that $\bbeta\ne 0$. Then, $-\hat{R}^*(\bbeta,Q)$ is a strictly quasi-convex function of $Q^t$, $t=1,2,\ldots,s$, for $Q\in \calQ'$.
\end{Lemma_A}
\begin{IEEEproof}
For each $Q\in\calQ'$, consider
\begin{align}
&\sum_{i=1}^n\sum_{j=1}^n\beta_{i}\beta_{j}g_{i}g_{j}\kappa_{Q}(\bx_i,\bx_j)\nonumber\\
&= \sum_{i,k}\sum_{j,k'}Q(\bz_k|\bx_i)Q(\bz_{k'}|\bx_j)\beta_{i}\beta_{j}g_{i}g_{j}\kappa(\bz_{k},\bz_{k'})\label{convex}\\
&=\norm{\sum_{i,k}\beta_{i}g_{i}Q(\bz_k|\bx_i)\Phi(\bz_k)}^2> 0,\nonumber
\end{align}
where the last inequality holds due to Assumption~\ref{assumpt:phi2bayes}\ref{it:strictly positive} since $\bbeta\neq\mathbf{0}$, and $Q^t\ne 0$ for all $t=1,\ldots,s$ because of \eqref{Qnsmall}.
From \eqref{dualR} and \eqref{convex}, $-\hat{R}^*(\bbeta,  Q)$ is a positive definite quadratic form of $Q^t$, for each $t=1,2,\ldots,s$, and is thus a strictly convex function of $Q^t$. This implies strict quasi-convexity, and the proof of the lemma is complete.
\end{IEEEproof}

We now show that  $\{\bbeta[k]\}$ and $\{Q[k]\}$ in Algorithm~\ref{algo:step1} have limit points. Since $\phi^*(\cdot) \geq -\phi(0) > -\infty$ , $-\hat{R}^*(\bbeta,  Q)$ is lower bounded. Therefore, the decreasing sequence $F[k]=-\hat{R}^*(\bbeta[k], Q[k])$ converges \cite{Rudin1964}. From Proposition 4 of \cite{Grippo2000}, convergence of $\{F[k]\}$ implies convergence of $\{\bbeta[k]\}$ and $\{ Q[k]\}$ to limit points. By Proposition~5 of \cite{Grippo2000}, this shows that Algorithm~\ref{algo:step1} converges to a critical point. The proof of the proposition is now complete.

\section{Proof of Proposition~\ref{pro:converge2}}\label{prf:converge2}

From  \cite{NguWaiJor:05}, $F(w_H,Q)$ is a convex function of $Q^t$, for each $t=1,2,\ldots,s$. Since $-\log(\cdot)$ is a non-increasing convex function, and $\hat{R}^*(\bbeta,Q)-\theta$ is a strictly concave function of $Q^t$, $-\frac{1}{\mu}\log(\hat{R}^*(\bbeta, Q)-\theta)$ is strictly convex with respect to $Q^t$ \cite{Boyd2004}. Therefore, $F_0(w_H,\bbeta,Q)$ is a strictly quasi-convex function of $Q^t$, $t=1,\ldots,s$ (see Definition~\ref{def:quasi-convex}). The rest of the proof is similar to that in Appendix~\ref{prf:converge1}.

\section{Proof of Theorem~\ref{thm:m-empricalInfoPrivacy}}\label{prf:m-empricalInfoPrivacy}

Following the same argument as that in the proof of Proposition~\ref{thm:PrivacyMetrics}\ref{it:infoprivacy-bayes}, for any $1\leq g < m$ and $\bz\in\calD'$, we have from \eqref{ineq:ellzbounds},
\begin{align*}
 e^{-\epsilon/2} \leq \ell_g(\bz) \leq  e^{\epsilon/2},
\end{align*}
and for any $g,g'\in \{0,\ldots,m-1\}$,
\begin{align*}
e^{-\epsilon} \leq \frac{p_{Z\mid G}(\bz\mid g)}{p_{Z\mid G}(\bz\mid g')} \leq  e^{\epsilon}.
\end{align*}
The proof then proceeds similarly as that in Proposition~\ref{thm:PrivacyMetrics}\ref{it:infoprivacy-bayes}.

\end{appendices}

\end{document}